\documentclass[10pt,letterpaper]{article}
\usepackage[top=0.85in,left=2.75in,footskip=0.75in]{geometry}

\usepackage{amsmath,amssymb}

\usepackage{changepage}

\usepackage[utf8x]{inputenc}

\usepackage{textcomp,marvosym}

\usepackage{cite}

\usepackage{nameref,hyperref}
\usepackage[dvipsnames]{xcolor}

\usepackage[right]{lineno}

\usepackage{microtype}
\DisableLigatures[f]{encoding = *, family = * }

\usepackage{array}
\usepackage{amsmath,amssymb}
\usepackage{multirow}
\usepackage{graphics,graphicx,subfigure}
\usepackage{dcolumn,bm}
\usepackage{psfrag,mathtools}
\usepackage{color}

\usepackage{psfrag}
\usepackage{gensymb}
\usepackage[dvipsnames]{xcolor}
\usepackage{ragged2e}

\newcolumntype{+}{!{\vrule width 2pt}}

\newlength\savedwidth

\raggedright
\usepackage[aboveskip=1pt,labelfont=bf,labelsep=period,justification=raggedright,singlelinecheck=off]{caption}

\makeatletter
\renewcommand{\@biblabel}[1]{\quad#1.}
\makeatother

\usepackage{lastpage,fancyhdr,graphicx}
\usepackage{epstopdf}

\fancyheadoffset[L]{2.25in}
\fancyfootoffset[L]{2.25in}
\begin{document}
\vspace*{0.2in}

\begin{flushleft}
{\Large
\textbf\newline{Analysis of emergent patterns in crossing flows of pedestrians reveals an invariant of `stripe' formation in human data}
}
\newline
\\
Pratik Mullick\textsuperscript{1*},
Sylvain Fontaine\textsuperscript{2\textcurrency},
Cécile Appert-Rolland\textsuperscript{2},
Anne-Hélène Olivier\textsuperscript{3},
William H. Warren\textsuperscript{4},
Julien Pettré\textsuperscript{1}
\\
\bigskip
\textbf{1} INRIA Rennes - Bretagne Atlantique, Campus de Beaulieu, Rennes, France
\\
\textbf{2} Université Paris-Saclay, CNRS/IN2P3, IJCLab, Orsay, France
\\
\textbf{3} Université Rennes, INRIA, CNRS, IRISA, M2S, France
\\
\textbf{4} Department of Cognitive, Linguistic and Psychological Sciences, Brown University, Providence, Rhode Island, USA
\\
\bigskip
\textcurrency Current Address: Sorbonne Université, CNRS - GEMASS, Paris, France

* pratik.mullick@inria.fr

\end{flushleft}

\section*{Abstract}
\justifying

When two streams of pedestrians cross at an angle, striped patterns spontaneously emerge as a result of local pedestrian interactions. This clear case of self-organized pattern formation remains to be elucidated. In counterflows, with a crossing angle of $180\degree$, alternating \textit{lanes} of traffic are commonly observed moving in opposite directions, whereas in crossing flows at an angle of $90\degree$, diagonal \textit{stripes} have been reported.  Naka (1977) hypothesized that stripe orientation is perpendicular to the bisector of the crossing angle.  However, studies of crossing flows at acute and obtuse angles remain underdeveloped.  We tested the bisector hypothesis in experiments on small groups (18-19 participants each) crossing at seven angles ($30\degree$ intervals), and analyzed the geometric properties of stripes.  We present two novel computational methods for analyzing striped patterns in pedestrian data: (i) an edge-cutting algorithm, which detects the dynamic formation of stripes and allows us to measure local properties of individual stripes; and (ii) a pattern-matching technique, based on the Gabor function, which allows us to estimate global properties (orientation and wavelength) of the striped pattern at a time $T$.  We find an invariant property: stripes in the two groups are parallel and perpendicular to the bisector at all crossing angles.  In contrast, other properties depend on the crossing angle:  stripe spacing (wavelength), stripe size (number of pedestrians per stripe), and crossing time all decrease as the crossing angle increases from $30\degree$ to $180\degree$, whereas the number of stripes increases with crossing angle.  We also observe that the width of individual stripes is dynamically squeezed as the two groups cross each other.  The findings thus support the bisector hypothesis at a wide range of crossing angles, although the theoretical reasons for this invariant remain unclear.  The present results provide empirical constraints on theoretical studies and computational models of crossing flows.

\section*{Author summary}

You may have noticed that pedestrians in a crosswalk often form multiple \textit{lanes} of traffic, moving in opposite directions ($180\degree$). Such spontaneous pattern formation is an example of self-organized collective behavior, a topic of intense interdisciplinary interest.  When two groups of pedestrians cross at an intersection ($90\degree$), similar diagonal \textit{stripes} appear.  Naka (1977) conjectured that the stripes are perpendicular to the mean walking direction of the two groups.  This facilitates the forward motion of each group and reduces collisions. We present the first empirical test of the hypothesis by studying two groups of participants crossing at seven different angles ($30\degree$ intervals).  To analyze the striped patterns, we introduce two computational methods, a local Edge-cutting algorithm and a global Pattern-matching technique.  We find that stripes are indeed perpendicular to the mean walking direction at all crossing angles, consistent with the hypothesis.  But other properties depend on the crossing angle: the number of stripes increases with crossing angle, whereas the spacing of stripes, the number of pedestrians per stripe, and the crossing time all decrease.
Moreover, the width of individual stripes is ``squeezed'' in the middle of the crossing.
Future models of crowd dynamics will need to capture these properties.

\section*{Introduction}
Collective motion in groups of humans, as well as other social organisms, has increasingly become a subject of analysis and modeling \cite{john2004,gautrais2012,mehdi2012,andrea2012,fujii2018,strombom2018,rio2018}. Currently, characteristic patterns of collective motion are understood as emergent behavior resulting from the collective dynamics of interactions between individuals.  Studies of human crowd dynamics have important applications to improving pedestrian traffic flow, safety management, and the prevention of crowd disasters \cite{thompson1995,helbingnature2000,sharma2001,yangchen2013,bohannon2005}. Analyses of real-life mass events have been used to model crowd behavior in situations such as religious gatherings, rock concerts, sporting matches, and transportation hubs \cite{hoogendorn2004, helbinghajj2007,  johansson2008, bain2018}, with a critical goal of averting life-threatening crushes, stampedes, and trampling \cite{helbinghajj2007,batty2003,lee2005}.
A first step to successful modeling is a better understanding of actual crowd behavior by analysis of crowd dynamics and pattern formation in human data.  In this paper, we develop a computational analysis of spontaneous stripe formation in crossing flows of pedestrians.

Pedestrian traffic flow has been studied empirically in a wide variety of situations, using both experimental methods and motion tracking of real crowds.  The simplest case is uni-directional flow in a corridor, in which properties such as the dependence of speed on density have been analyzed  \cite{seyfried2005,chattaraj2009,seyfried_book2010,jezbera2010,yana2012,asja2012}.
Collision avoidance between pedestrians has been investigated in pairs of walkers \cite{oliver2012,oliver2013}
and multiple walkers
\cite{laurentius2018}.
Bottlenecks occur when a large group attempts to pass through a narrow opening \cite{seyfried_book2010,hoog2005, kretz2006_1,seyfried2009,faure2015,garci2016,nicolas2017},
as in Black Friday sales or fire emergencies, which can lead to jamming and crushes.  Other empirical studies have examined pedestrian flow through a T-junction 
\cite{zhang2011,zhang2013},
multi-directional flows
\cite{cao2017},
a pedestrian crossing through a dense static crowd 
\cite{cecile2018,nicolas2019},
and a bottleneck leading to a 1D corridor
\cite{adrian2020}.

\textit{Crossing flows} can be described as two streams of pedestrians walking in different directions, passing through each other at a crossing angle $\alpha>0\degree$ (where $0\degree$ is walking in the same direction).  Many real-world situations produce crossing flows, such as streams of pedestrians crossing at a sidewalk intersection, or subway commuters passing each other when entering and exiting a metro car. A special case of crossing flows, called \textit{counterflow}, occurs when the crossing angle is $180\degree$.  Self-organized spatial patterns have been observed when two groups cross each other. In counterflow, the formation of stable \textit{lanes} is regularly reported in both human experiments  \cite{daamen2004,helbing2005,kretz2006,mehdi2012,zhang2012,feliciani2016,cao2017}
and numerical simulations 
\cite{socialforce,fukui1999,muramatsu1999,helbingrmp2001,burstedde2001,blue2001,tajima2002,dzubiella2002,hoogendoorn_b2003,isobe2004,nagai2005,cristin2019,khelfa2021}, 
in which alternating lanes of pedestrian traffic are aligned with the walking directions of the two groups ($180\degree$ apart). A jamming transition can occur above a critical flow density \cite{muramatsu1999,tajima2002,isobe2004,nagai2005}.
More generally, at other crossing angles the formation of stripes is observed, but the alternating stripes are not aligned with the walking directions of the two groups.  The familiar case of orthogonal flows ($\alpha=90\degree$) has been widely studied, and the formation of diagonal stripes is found in human crowds \cite{hoogendoorn_b2003,naka1977} and in simulation \cite{dzubiella2002,hoogendoorn_b2003,daamen2004,ondrej2010b,yamamoto_o2011,cividini1,cividini2,cividini3,cividini4}.
However, the analysis of crossing flows in humans at other crossing angles remains underdeveloped.

Naka \cite{naka1977} first reported stripes at acute and obtuse crossing angles in pedestrian crowds, and hypothesized that stripes form at an orientation that is perpendicular to the bisector of the crossing angle. Abstractly, a stripe is a traveling wave that moves in the mean direction of the two flows, such that individual pedestrians travel forward with a stripe and laterally within it \cite{helbing2005}.  A striped pattern facilitates overall pedestrian flow by reducing collision-avoidance maneuvers, thereby increasing the average walking speed.  Only one subsequent human study \cite{wong2010} has tested oblique crossing angles (45$\degree$ and 135$\degree$), but stripe patterns were not analyzed.  The bisector hypothesis thus remains to be tested experimentally. 

Striped patterns in oblique crossing flows have been reproduced in simulation, consistent with the bisector hypothesis \cite{helbing2005}. In one system, the inclination of the stripe to the bisector was found to increase with the velocity difference between two orthogonal flows \cite{dzubiella2002}. The mechanism responsible for the formation of self-organized stripes in orthogonal flows has been studied theoretically \cite{cividini1,cividini2,cividini3,cividini4}. A mean field analysis shows the underlying mechanism to be a linear instability of the randomly uniform state in the intersecting region compared to the formation of diagonal striped patterns \cite{cividini1,cividini3,cividini4}.  The `wake' of a pedestrian has been proposed as the microscopic mechanism for stripe formation, a density perturbation created in the perpendicularly moving flow \cite{cividini2}. The inclination of the striped patterns was related to the velocity difference between the two groups, producing a `chevron' effect \cite{cividini1,cividini3}. Absence of striped patterns has also been observed when three or more groups of people intersect \cite{helbing1997,helbingmolnar1997,helbingvicsek1999}.

The purpose of the present research is to experimentally test the bisector hypothesis by analyzing stripe formation at a variety of crossing angles, without spatial constraints.  We seek to answer several theoretically-motivated questions: (i) Can stripe orientation be predicted as perpendicular to the bisector for all crossing angles?  (ii) Do other stripe properties depend on crossing angle?  (iii) What are the stripe dynamics during crossing flows?  (iv) Does spontaneous stripe formation generalize from continuous crossing flows in defined corridors to small crowds without boundary conditions on spatial position, density, or visibility?

\begin{figure}[h!]
\begin{adjustwidth}{-2.0in}{0in}
    \centering
\includegraphics[width=17.5cm]{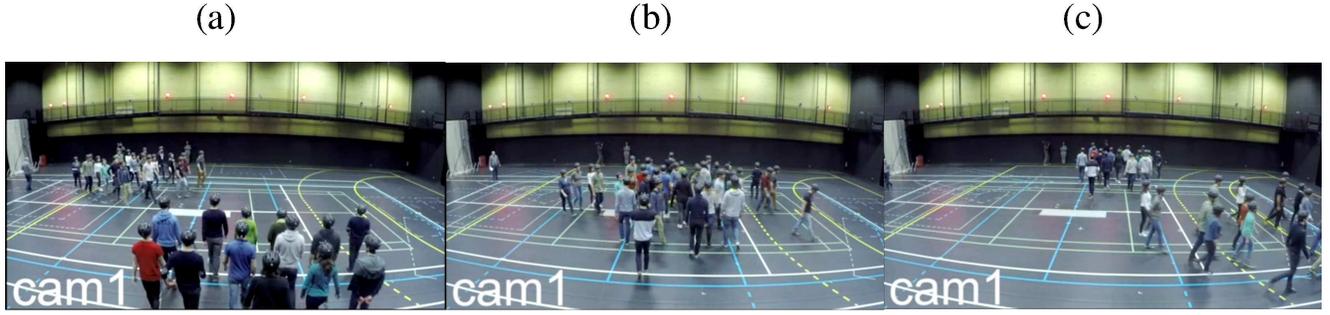}
    \vspace{0.3cm}
    \caption{\textbf{Photograph of our experimental set-up to study crossing flows.} Agents participating in our experiment are shown in this photograph for a typical trial with crossing angle $120\degree$. The three stages of the trial are shown here, viz. (a) before crossing (b) during crossing and (c) after crossing.}
    \label{real_photo}
\end{adjustwidth}
\end{figure}

\begin{figure}[h!]
\begin{adjustwidth}{-2.0in}{0in}
    \centering
     \includegraphics[width=18cm]{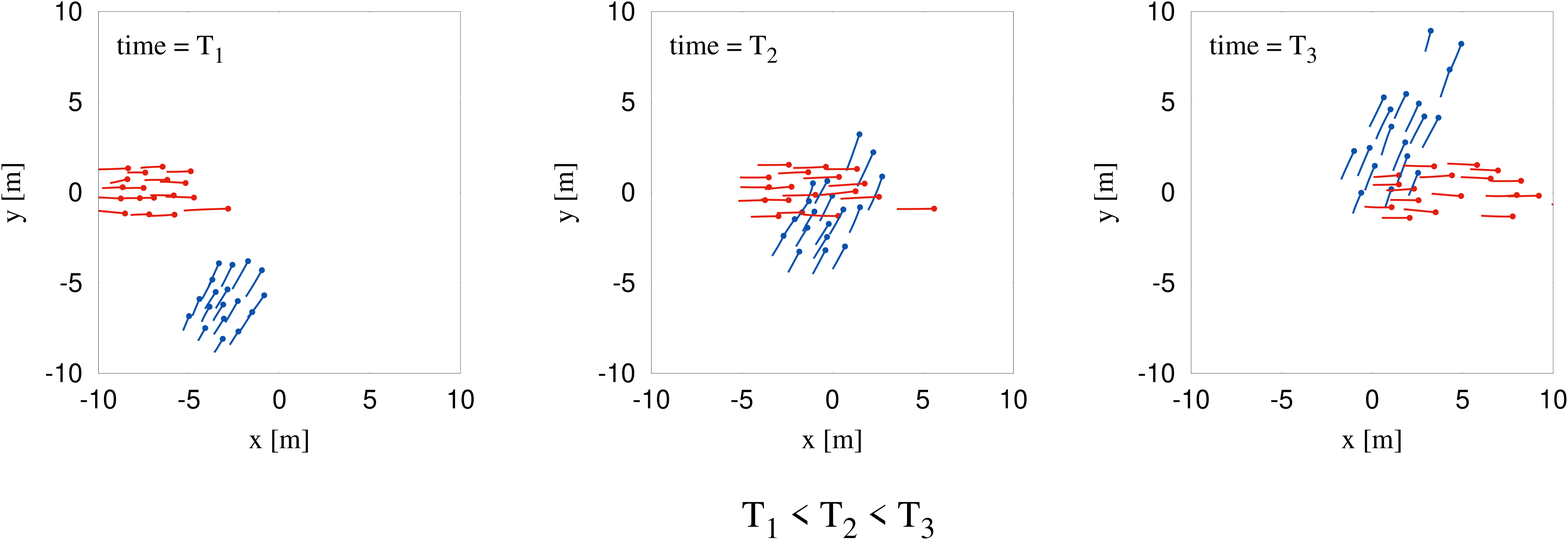}
     \vspace{0.1cm}
    \caption{\textbf{Illustration of a trial of crossing flow from our experiments.} Traces of all the pedestrians involved for a typical trial has been shown with expected value of crossing angle equal to $60\degree$. Three different instances of the trial has been shown here viz. before crossing ($T_1$), during crossing ($T_2$) and after crossing ($T_3$). The actual values of time frames are $T_1=2.3$ sec, $T_2=6.55$ sec and $T_3=10.8$ sec from the beginning of the trial. The two groups of pedestrians are denoted by blue and red dots. The tails behind each of the dots are basically the distances travelled by the pedestrians in previous 1.25 sec.}
    \label{traj}
\end{adjustwidth}
\end{figure}

We addressed these questions as part of the PEDINTERACT Project \cite{cecile2018}, in which two different sets of subjects participated (36 on Day 1, 38 on Day 2).  The setup appears in Fig \ref{real_photo} (also see \nameref{S1_video}). In the experiment, two groups of participants (18 or 19 per group) walked through each other at seven different crossing angles ($0\degree$ to $180\degree$, at $30\degree$ intervals); there were approximately 17 trials per angle. On each trial, the groups were positioned in two starting boxes oriented at the designated crossing angle, and were instructed to walk in the direction they were facing to the other side of the room. To investigate whether striped patterns would emerge in the absence of spatial boundary conditions, we did not use opaque corridors as in many previous studies \cite{helbing2005,ondrej2010b,cividini1,wong2010}. Head position was recorded with a motion-capture system at 120 Hz. Sample traces for all pedestrians in a typical trial appear in Fig \ref{traj}.

Because the empirical analysis of crossing flows is quite underdeveloped, we describe a number of computational methods for analyzing the characteristics of stripes in human data.  In particular, we present a novel approach to identify the formation of stripes, called the Edge-cutting algorithm. Using this algorithm we were able to measure the local properties of individual stripes such as their orientation, width and size. We also use an independent method to characterize global stripe properties, a pattern matching technique that fits a two dimensional sinusoidal function (e.g. Gabor function) to the positions of pedestrians in the two groups. This method assumes the existence of a periodic pattern of stripes and then finds the geometric properties of the pattern from a fitting procedure. The two methods are complementary, in the sense that the edge-cutting algorithm requires the whole history of the crossing and provides the full dynamics of the stripes, whereas the pattern matching method can be performed on a single snapshot. The stripe orientations obtained by the edge-cutting algorithm and the pattern matching technique are compared to each other, and to the hypothesis that the stripes are perpendicular to the bisector of the crossing angle.

In sum, the present paper makes two major contributions: (i) we present experimental data on crossing flows of pedestrians that support the bisector hypothesis, and (ii) we introduce and discuss methodological tools to detect the formation and presence of striped patterns and to estimate their geometric properties.

\section*{Results}

\begin{figure}[h!]
\begin{adjustwidth}{-2.0in}{0in}
    \centering
    \includegraphics[width=18cm]{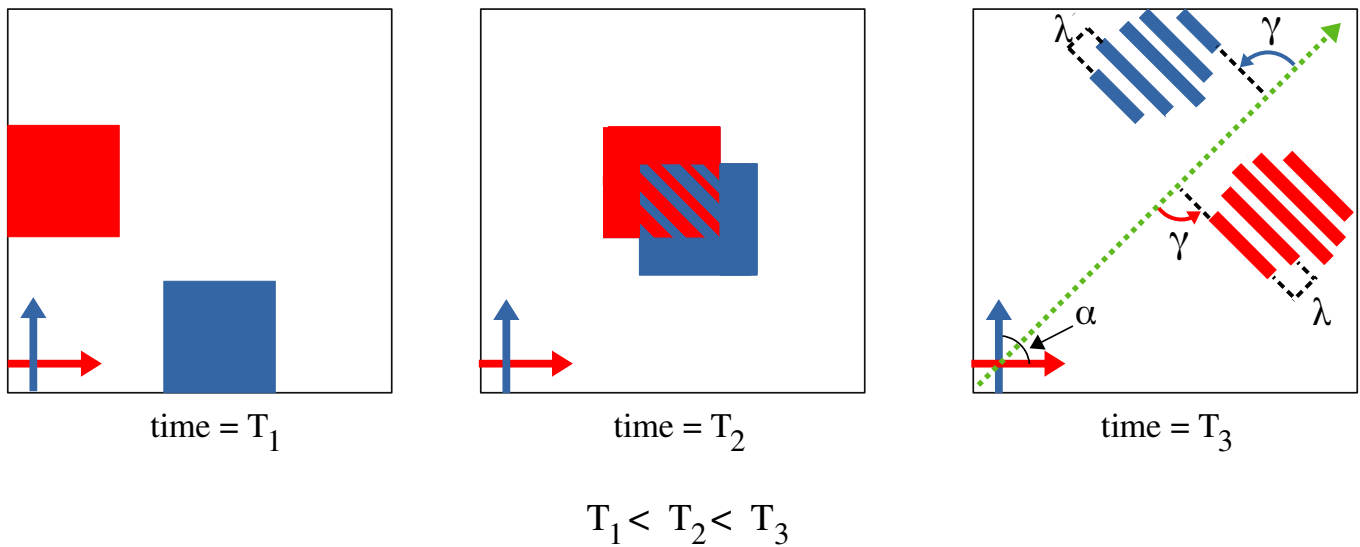}
    \vspace{0.1cm}
    \caption{\textbf{Schematic representation for the formation of stripes and definition of orientation $\gamma$ and physical separation $\lambda$ of stripes.} Formation of stripes as a consequence of two groups crossing each other. In this schematic diagram the crossing angle between the two groups is $\alpha$. The figure has been shown for three instances viz. before crossing ($T_1$), during crossing ($T_2$) and after crossing ($T_3$). The two groups before crossing are denoted by blue and red squares, whose direction of motion is denoted by arrows of the same color. The green dotted arrow denotes the bisector of the crossing angle. The orientation $\gamma$ of the stripes is measured counter-clockwise from the bisector. $\lambda$ is the spatial separation between two stripes from the same group. For specific definitions of $\gamma$ and $\lambda$ see Fig \ref{schem_3M}.}
    \label{schem_str_form}
\end{adjustwidth}
\end{figure}

When two groups of people cross each other, striped patterns emerge, as schematically illustrated in Fig \ref{schem_str_form}. The primary goal of the present research was to characterize the properties of these emergent stripes, based on numerical analysis of participant trajectories.  The actual crossing angle $\alpha$ between the mean walking directions of the two groups was measured from the data.  The properties of stripe orientation $\gamma$ relative to the crossing angle bisector, and stripe spacing $\lambda$ are illustrated in Fig \ref{schem_str_form} (right).  We begin by introducing two independent computational methods devised to analyze the geometric properties of the stripes, (i) the Edge-cutting algorithm and (ii) the Pattern-matching technique.

\subsection*{Identifying stripes using edge-cutting algorithm}

\begin{figure}[h!]
\begin{adjustwidth}{-2.0in}{0in}
    \centering
    \includegraphics[width=18cm]{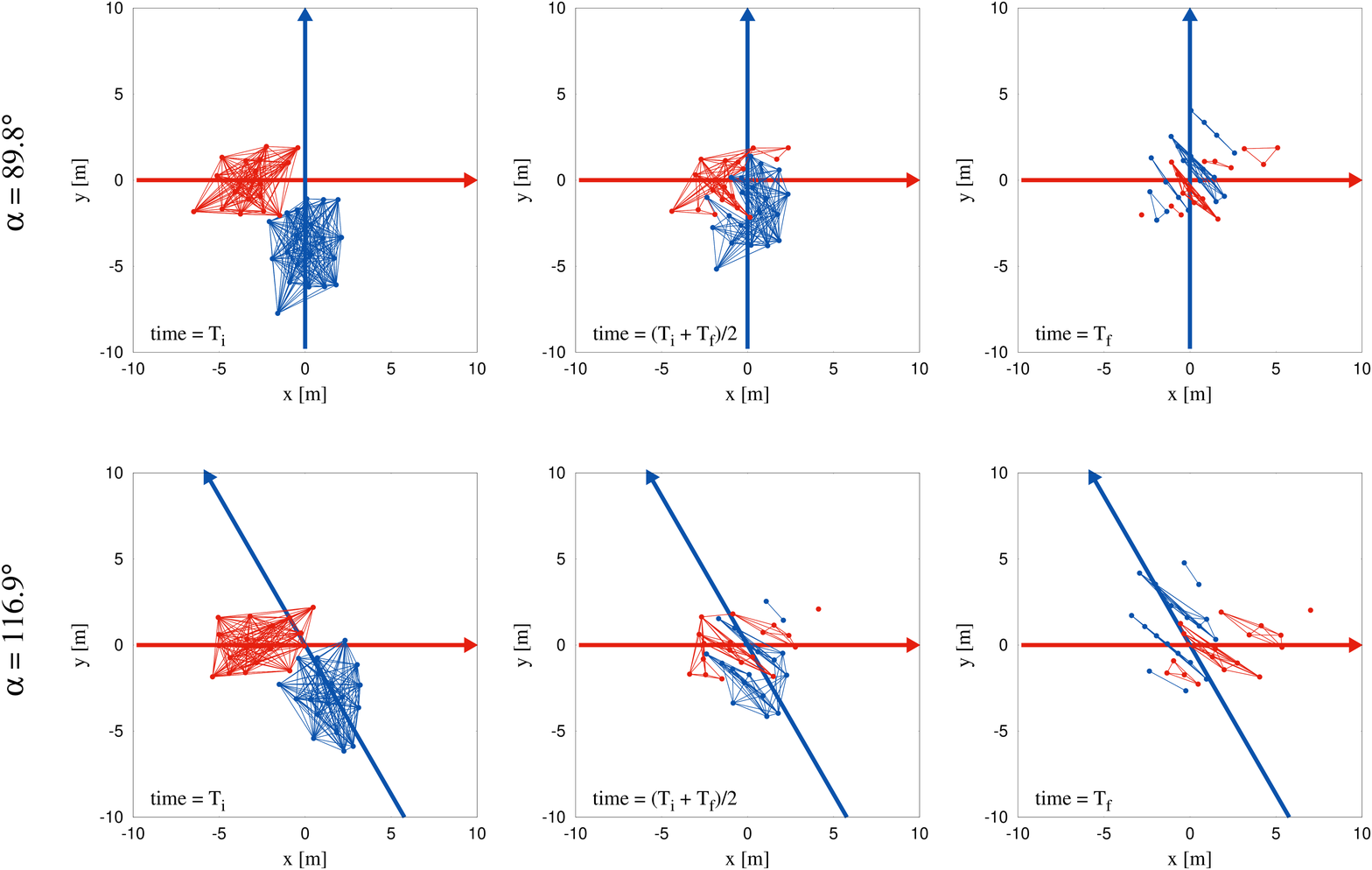}
    \vspace{0.1cm}
    \caption{\textbf{Pictorial representation of the edge-cutting algorithm.} Figure demonstrates the working process of the edge-cutting algorithm as a sequence of time. Here we show the process for two typical trials with $\alpha = 89.8\degree$ and $\alpha = 116.9\degree$. Red and blue arrows indicate the direction of motion of the two groups represented by red and blue dots respectively. The lines connecting the dots in each of the groups are considered as the virtual bonds or `edges' 
    which are suppressed when cut by a pedestrian on the other group
    (see Materials and Methods). The figures are shown for three instances, viz. $T_i$, $(T_i+T_f)/2$ and $T_f$. $T_i$ and $T_f$ denote the instances of time when the first and last edge-cut take place respectively. The edge-cutting process for the entire course of time for these two trials are shown as videos in supplementary materials (\nameref{S2_video} and \nameref{S3_video}).}
    \label{edge_ex}
\end{adjustwidth}    
\end{figure}

For purposes of the first method, we define a stripe as a subset of participants from one group that is not penetrated by participants from the other group. Specifically, the virtual connections or edges between the participants in a stripe are never crossed or `cut' by the trajectory of a participant from the other group. The principal output of the Edge-cutting algorithm is the identification of the participants who belong to each stripe (see Materials and Method for details). This analysis indeed revealed the spontaneous emergence of striped patterns and the stripes were successfully identified. The dynamics of stripe formation was also observed, as illustrated for two typical trials in Fig \ref{edge_ex}. The Edge-cutting algorithm also yields the time of the initial edge-cut $T_i$ at the start of crossing (left column of Fig \ref{edge_ex}) and the time of the final edge-cut $T_f$ at the end of crossing (right column of Fig \ref{edge_ex}). Animations of the Edge-cutting process for these two trials appear in the supplementary material (\nameref{S2_video} and \nameref{S3_video}).

\subsection*{Characterizing stripes using pattern-matching technique}

\begin{figure}[h!]
\begin{adjustwidth}{-2.0in}{0in}
    \centering
    \includegraphics[width=15cm]{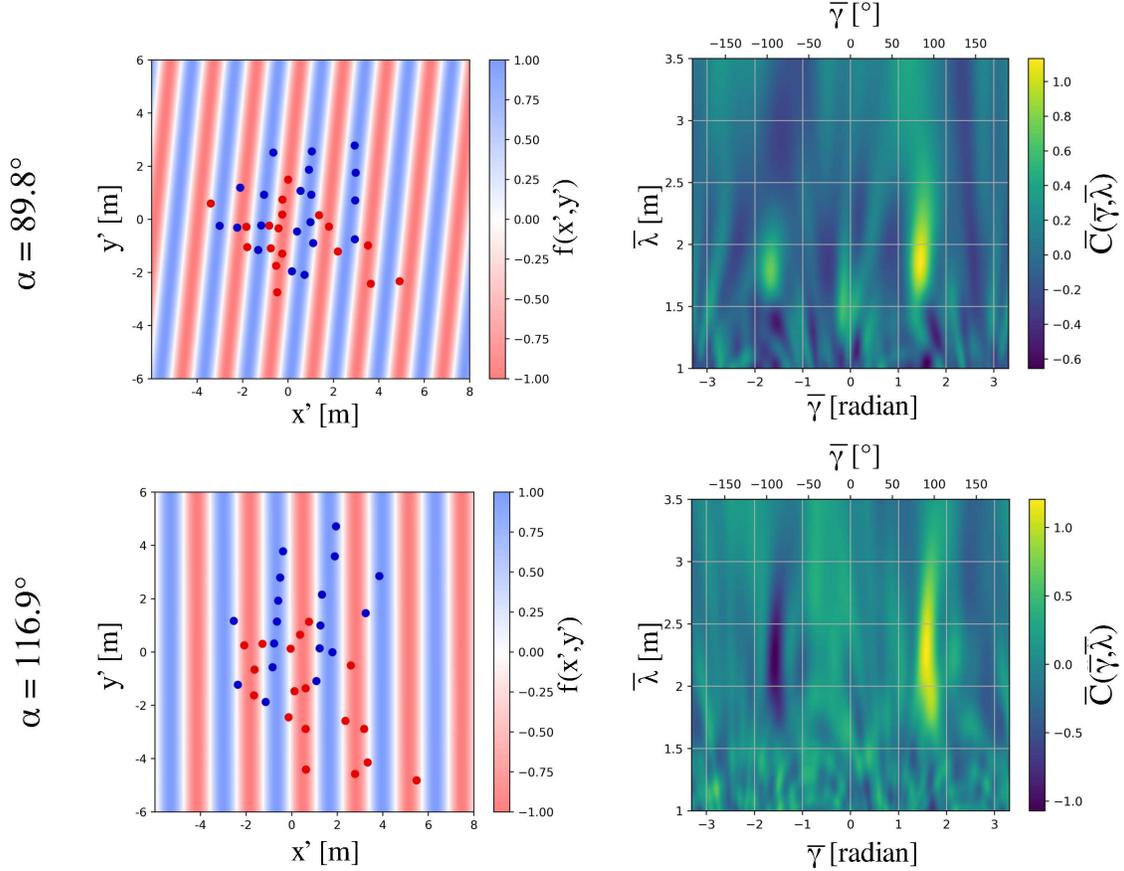}
    \vspace{0.3cm}
    \caption{\textbf{Output of the pattern matching technique.} The panels on the left show fitting of pedestrian positions for the same trials shown in Fig \ref{edge_ex}. Fitting was done using the parametric sine curve $f$ and the transformed coordinates $(x',y')$. The blue and red parts of the plot represent the crests and troughs of the sine function respectively.
    The outputs of the fitting are $\Bar{\gamma}=83.34\degree$, $\Bar{\lambda}=1.865$m and $\Bar{\psi}=-116.37\degree$ for the typical trial with $\alpha=89.8\degree$ and $\Bar{\gamma}=89.99\degree$, $\Bar{\lambda}=2.286$m and $\Bar{\psi}=-0.92\degree$ for the typical trial with $\alpha=116.9\degree$. (see Materials and Methods) The panels on the right show variation of $\Bar{C}$ as a function of $\Bar{\gamma}$ and $\Bar{\lambda}$ keeping $\psi$ fixed to the value obtained from the fitting shown in the left panel. The region of occurrence of high values of $\Bar{C}$ is shown in yellow. The function $\Bar{C}$ was maximised to fit the sine function $f$. The maximum value of $\Bar{C}$ for the trial with $\alpha=89.8$, as obtained by our optimisation procedure is 1.132, which occurred for $\Bar{\gamma}=83.34\degree$ and $\Bar{\lambda}=1.865$m. Whereas, for the trial with $\alpha=116.9\degree$ we obtained the maximum value of $\Bar{C}$ = 1.205, which occurred for $\Bar{\gamma}=89.99\degree$ and $\Bar{\lambda}=2.286$m.}
    \label{edge_gabor}
\end{adjustwidth}    
\end{figure}

The Pattern-matching technique estimates the orientation and width of a set of stripes, assuming that the stripes are parallel and equally spaced.  This method fits a two-dimensional spatial frequency function $f$, based on a sinusoidal Gabor function, to the positions of pedestrians  at a time $T$. The free parameters of orientation $\gamma$, wavelength $\lambda$, and phase $\psi$, are chosen by maximizing $C$, the fit of the function to pedestrian positions, where positive values (peaks) are assigned to one group and negative values (troughs) to the other (see Materials and Methods for details). The fitting can be applied to all pedestrians or to a subset (e.g. one group). The output of this fitting procedure for all pedestrians in two representative trials appears in Fig \ref{edge_gabor}, where $\Bar{\gamma}$ and $\Bar{\lambda}$ refer to the orientation and wavelength of stripes in the whole crowd.

The two methods are complementary. The edge-cutting algorithm requires the whole history of the crossing event and yields in return the full dynamics of stripes. It estimates the local spatial properties of individual stripes at each time point, without assuming any prior knowledge about them. The pattern-matching technique requires only a single snapshot, and recovers the global spatial properties of orientation and wavelength. The prior assumptions of parallel, equally spaced stripes help to match the instantaneous pattern, as long as the actual stripes are close to this ideal.  We will see now how both approaches allow us to gain insight into the striped structure.

\subsection*{Stripe Orientation}

\begin{figure}[h!]
\begin{adjustwidth}{-2.0in}{0in}
    \centering
    \includegraphics[width=18cm]{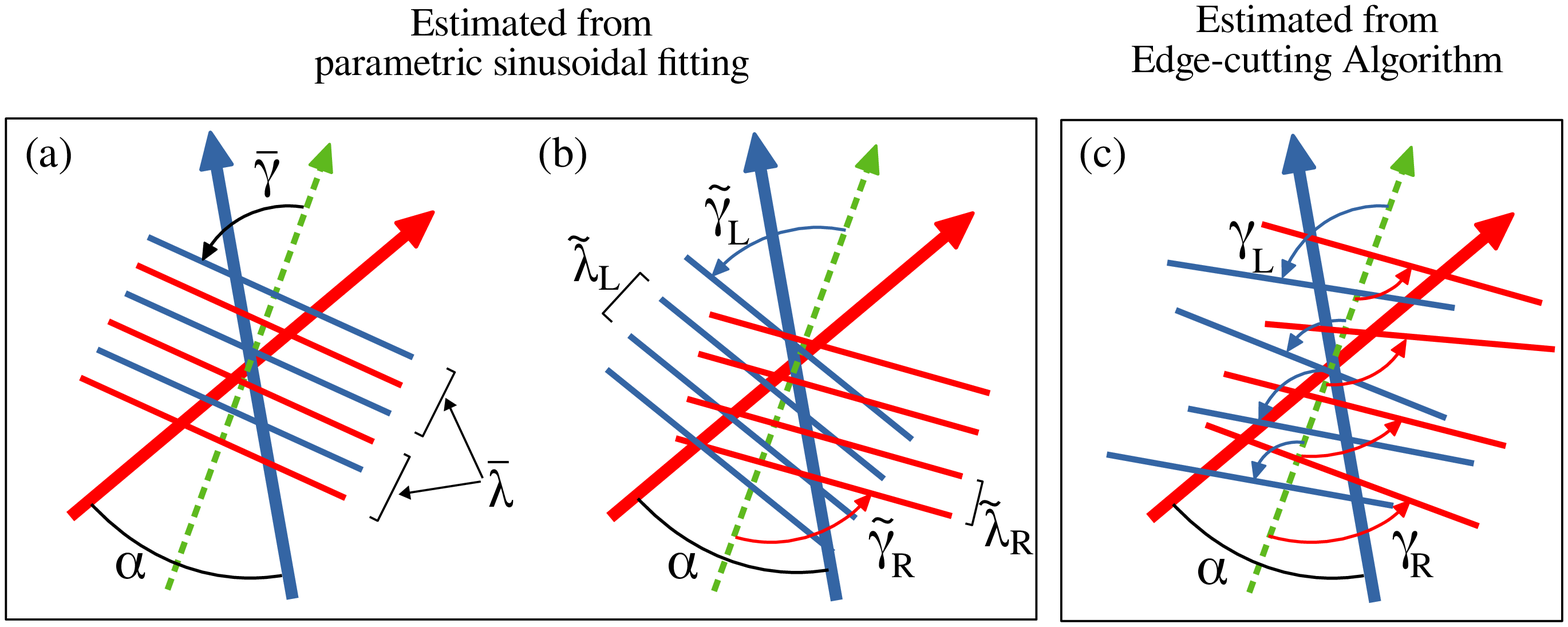}
    \vspace{0.5cm}
    \caption{\textbf{Summary of different methods to estimate orientation $\gamma$ and physical separation $\lambda$ of the stripes.} The arrows in blue and red represents the direction of motion of the two groups. The schematic diagrams are shown for an arbitrary crossing angle. The dashed green arrow indicates the bisector of the crossing angle between the two group direction vectors. The lines in blue and red show the stripes from the two groups. $\gamma$ is the angle between the direction of stripes and the bisector of crossing angle, always measured counterclockwise. (a) Estimation of orientation $\Bar{\gamma}$ of the stripes and physical separation $\Bar{\lambda}$ between two stripes from the same group using the parametric sinusoidal fitting. In doing this calculation it was assumed that stripes from the two groups are parallel to each other and are equispaced, as shown in the figure. (b) Orientation of the stripes from the two groups when we assume that stripes from the same group are parallel to each other and are equispaced. $\Tilde{\gamma}_L$ and $\Tilde{\gamma}_R$ denote the orientation of stripes whose group direction vectors are left and right to the direction of bisector respectively. Using the same convention, $\Tilde{\gamma}_L$ and $\Tilde{\gamma}_R$ are the spatial separation
    between the stripes in those cases. This calculation was also done by fitting the two dimensional sine curve. (c) Estimation of orientation of the individual stripes that were found using the edge-cutting algorithm, for the two groups.  
    $\gamma_L$ or $\gamma_R$ denote the orientation of individual stripes whose group direction vector is left or right to the direction of bisector respectively.}
    \label{schem_3M}
\end{adjustwidth}
\end{figure}

Based on previous empirical observations and modeling, there are reasons to expect that the observed stripes would be parallel and perpendicular to the bisector of the crossing angle, as illustrated in Fig \ref{schem_str_form}.  This bisector hypothesis thus predicts $\gamma = 90\degree$ for all crossing angles $\alpha$.  However, it is possible that the stripes for one group (blue in Fig \ref{schem_str_form}) are not parallel to those for the other group (red), such that $\gamma_{\text{blue}}\neq \gamma_{\text{red}}$; also that individual stripes within a group are not parallel. We thus estimated the orientation of stripes using the global and local methods: (i) the Pattern-matching technique allowed us to estimate the overall stripe orientation $\Bar{\gamma}$ (Fig \ref{schem_3M}a), as well as the orientation for each group separately $\Tilde{\gamma}_L$ and $\Tilde{\gamma}_L$, (Fig \ref{schem_3M}b); (ii) the Edge-cutting algorithm enabled us to estimate the orientation of individual stripes $\gamma_L$ and $\gamma_R$, (Fig \ref{schem_3M}c).  We report each of these measurements of stripe orientation in turn.

\subsubsection*{Global stripe orientation $\Bar{\gamma}$ and $\Tilde{\gamma}$ using the Pattern-matching technique}

\begin{figure}
\begin{adjustwidth}{-2.0in}{0in}
    \centering
    \includegraphics[width=18cm]{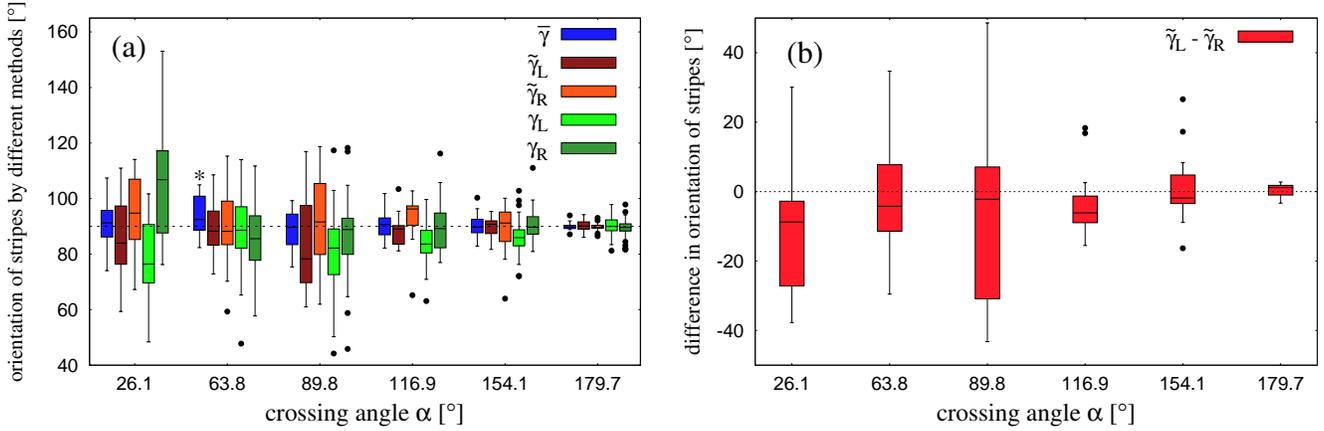}
    \vspace{0.5cm}
    \caption{\textbf{Orientation $\gamma$ of the stripes using different methods.} (a) Figure shows boxplots for obtained values of $\Bar{\gamma}$, $\Tilde{\gamma}_L$, $\Tilde{\gamma}_R$ from the pattern-matching technique and $\gamma_L$, $\gamma_R$ from the edge-cutting algorithm. Outliers are shown by black dots. The dashed line corresponds to $90\degree$. The boxplots were made using the various values of $\gamma$ evaluated at the same time instant. This instant was chosen to be the time when the periodicity of the stripes from the two groups was best maintained (see Discussion). For detailed definitions of $\Bar{\gamma}$, $\Tilde{\gamma}_L$, $\Tilde{\gamma}_R$, $\gamma_L$ and $\gamma_R$ see Fig \ref{schem_3M} and Materials and methods. (b) Boxplots for the difference in obtained values of orientations $(\Tilde{\gamma}_L - \Tilde{\gamma}_R)$ as estimated by separate-group analysis using pattern-matching technique. The values of $\Tilde{\gamma}_L$ and $\Tilde{\gamma}_R$ are the instantaneous values, which are shown in (a) (brown and orange bars)}
    \label{gamma_comp}
\end{adjustwidth}    
\end{figure}

In the first analysis, we estimated the overall orientation of all stripes $\Bar{\gamma}$ to the bisector, on the assumption that the stripes for the two groups were parallel and equally spaced (see Fig \ref{schem_3M}a), using the pattern matching technique.  The analysis was performed at a suitable time $T$ between the crossing midpoint $(T_i+T_f)/2$ and the final crossing point $T_f$, when the periodic pattern of stripes was most clearly defined (see Discussion).  The resulting values of $\Bar{\gamma}$ are represented in box plots in Fig \ref{gamma_comp}a (blue bars).  Note that the median values are very close to the predicted angle of $90\degree$ for all crossing angles, with deviations less than $3\degree$ in all conditions. A set of $t$-tests comparing the mean value of $\Bar{\gamma}$ to $90\degree$ at each crossing angle was significant only for $\alpha=63.8\degree$, $t(17)=-2.550$, $p=0.0207$; no other conditions were significant. Overall, this finding is consistent with the bisector hypothesis. A one-way analysis of variance (ANOVA) on $\Bar{\gamma}$ was also performed, with crossing angle $\alpha$ as the factor (excluding the $0\degree$ condition).  The result found that stripe orientation did not depend significantly on crossing angle, $F(5, 99) = 2.301$, $p = 0.0504$, $\eta^2 = 0.1$.

The second analysis added one degree of freedom by estimating the orientation of the stripes to the bisector for each group separately $(\Tilde{\gamma}_L, \Tilde{\gamma}_R)$, on the weaker assumption that only the stripes within one group are parallel.  We thus fit the sinusoidal function $f$ to the pedestrian positions separately for the two groups, yielding estimates of stripe orientation $\Tilde{\gamma}_L$ for the group heading to the left of the bisector, and $\Tilde{\gamma}_R$ for the group heading to the right (Fig \ref{schem_3M}b).  We first estimated these values at the same instant that the overall orientation $\Bar{\gamma}$ was estimated, and the results appear in the boxplots in Fig \ref{gamma_comp}a (brown and orange bars).  The median values are again close to $90\degree$ for all but one crossing angle, although the variability is greater because only half the pedestrian data contributed to each fit.  One-way ANOVAs found no influence of crossing angle on $\Tilde{\gamma}_L$, $F(5, 100) = 0.614$, $p = 0.689$, $\eta^2 = 0.029$, or on $\Tilde{\gamma}_R$, $F(5, 100) = 0.521$, $p = 0.76$, $\eta^2 = 0.025$. The within-trial difference between $\Tilde{\gamma}_L$ and $\Tilde{\gamma}_R$ is represented in Fig \ref{gamma_comp}b. The median values at each crossing angle lie close to $0\degree$, and the deviations are less than $5\degree$. A set of $t$-tests comparing $\Tilde{\gamma}_L - \Tilde{\gamma}_R$ to $0\degree$ at each crossing angle found no significant differences (all $p$’s $> 0.1$), indicating that the stripes for the left and right groups tend to be parallel within a trial.  

\begin{figure}[h!]
\begin{adjustwidth}{-2.0in}{0in}
    \centering
    \includegraphics[width=18cm]{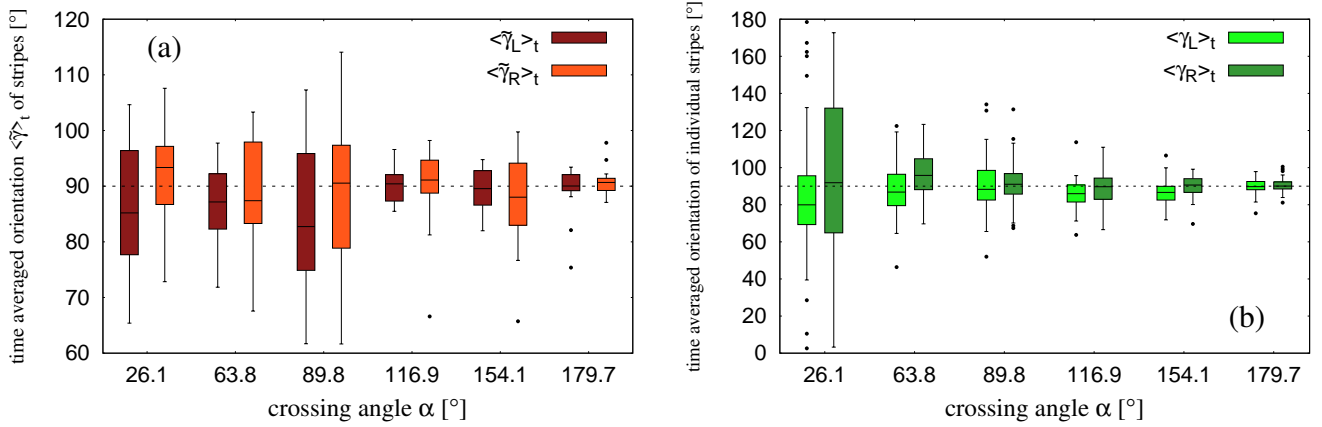}
    \vspace{0.3cm}
    \caption{\textbf{Time averaged 
    orientation of stripes.} (a) Boxplots for $\langle\Tilde{\gamma}_L\rangle_t$ and $\langle\Tilde{\gamma}_R\rangle_t$. $\Tilde{\gamma}_L$ and $\Tilde{\gamma}_R$ are the orientations of stripes in a trial as obtained from separate-group analysis using pattern matching technique. (b) Boxplots for $\langle\gamma_L\rangle_t$ and $\langle\gamma_R\rangle_t$, where $\gamma_L$ and $\gamma_R$ are estimated from the per-stripe analysis using edge-cutting algorithm.}
    \label{theta_avg}
\end{adjustwidth}    
\end{figure}

We then calculated $\Tilde{\gamma}_L$ and $\Tilde{\gamma}_R$ as a function of time, and obtained the time-average of these quantities, $\langle\Tilde{\gamma}_L\rangle_t$ and $\langle\Tilde{\gamma}_R\rangle_t$, during the interval $(T_i+T_f)/2$ to $T_f$ for each time frame of our data (see Discussion).  Boxplots of the time-averaged values appear in Fig \ref{theta_avg}a. Median angles are again close to the expected value of $90\degree$ for both left and right groups, with deviations less than $8\degree$ for acute crossing angles and less than $2\degree$ for obtuse angles. One-way ANOVAs did not find a significant effect of crossing angle $\alpha$ on either $\langle\Tilde{\gamma}_L\rangle_t$, $F(5, 100) = 0.934$, $p = 0.462$, $\eta^2 = 0.045$, or $\langle\Tilde{\gamma}_R\rangle_t$, $F(5, 100) = 0.399$, $p = 0.848$, $\eta^2 = 0.019$. A set of $t$-tests comparing $\langle\Tilde{\gamma}_L - \Tilde{\gamma}_R\rangle_t$ to $0\degree$ at each crossing angle found no significant differences (all $p$’s $> 0.16$). Together, these results indicate that the stripes formed by the two groups are generally parallel and perpendicular to the bisector of the crossing angle, as measured by the pattern-matching technique.

\subsubsection*{Individual stripe orientation ($\gamma_L$ and $\gamma_R$) based on the Edge-cutting algorithm}

Finally, we performed an analysis of each individual stripe based on the output of the Edge-cutting algorithm.  The orientations of stripes in the left and right groups are denoted $\gamma_L$ and $\gamma_R$ , respectively (see Fig \ref{schem_3M}c).  We first measured these values at the same instant that the overall orientation $\Bar{\gamma}$ was estimated, and the results appear in the boxplots in Fig \ref{gamma_comp}a (light-green and dark-green bars).  The medians for individual stripes are again close to the expected value of $90\degree$ and comparable to the other techniques, with the exception of the most acute crossing angle ($\alpha=26.1\degree$).

Because the configuration of stripes changed over time (e.g. Fig \ref{edge_ex}), we also measured the individual stripe orientations at different time points and computed time-averaged values, $\langle\gamma_L\rangle_t$ and $\langle\gamma_R\rangle_t$, during the interval $(T_i+T_f)/2$ to $T_f$ for each time frame of our data. The results of this analysis appear in the boxplots in Fig \ref{theta_avg}b.  Although there is much more variability in individual stripe orientations than in the pattern-matching estimates (Fig \ref{gamma_comp}a), the median values for both left and right groups are again quite close to the expected value of $90\degree$, consistent with the bisector hypothesis.

\subsection*{Other stripe properties}

We also used the global and local methods to estimate other geometric properties of the observed stripes, including their physical spacing, width, number, and size (number of pedestrians per stripe).

\subsubsection*{Stripe spacing ($\lambda$) using the Pattern-matching technique}

\begin{figure}[h!]
\begin{adjustwidth}{-2.0in}{0in}
    \centering
    \includegraphics[width=18cm]{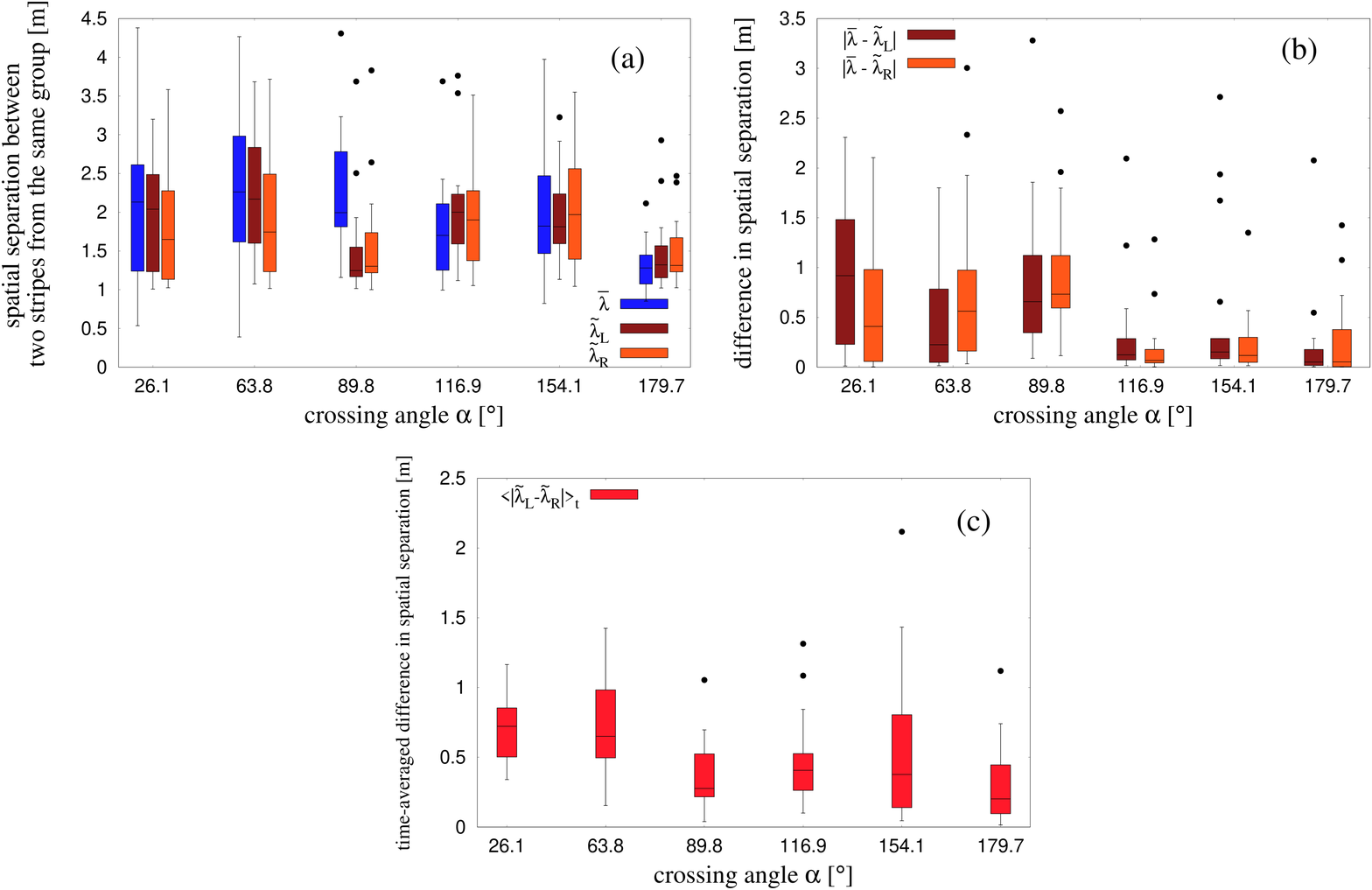}
    \vspace{0.5cm}
    \caption{\textbf{Boxplots for $\lambda$ as estimated from the pattern matching technique.} $\lambda$ is the spatial separation between two alternate stripes from the same group. (a) The boxplots were made over all the trials for obtained values of $\Bar{\lambda}$, $\Tilde{\lambda}_L$ and $\Tilde{\lambda}_R$ at the instant when the periodicity of the two groups was best maintained. $\Bar{\lambda}$, $\Tilde{\lambda}_L$ and $\Tilde{\lambda}_R$ are defined in Materials and methods. (b) Boxplots for the difference between obtained values of spatial separations from the whole-crowd and separate-group analyses under the pattern matching procedure. (c) Time-averaged difference of physical separation $\langle|\Tilde{\lambda}_L-\Tilde{\lambda}_R|\rangle_t$, where $\Tilde{\lambda}_L$ and $\Tilde{\lambda}_R$ are the physical separations between stripes in a trial as estimated from the separate-group analysis using pattern matching technique.}
    \label{pattern_output}
\end{adjustwidth}    
\end{figure}

We used the Pattern-matching technique to estimate the spacing of stripes, on the assumption that the stripes in the analysis are parallel and equally spaced.  The physical separation between the centers of stripes corresponding to the same group was estimated by the wavelength $\lambda$ of the sinusoidal function $f$, which was fitted to the pedestrian position data as described in the previous section.  The first analysis fit the data from the whole crowd to obtain $\Bar{\lambda}$, the overall spacing between stripes from the same group (see Fig \ref{schem_3M}a), at the same instant that the overall orientation $\Bar{\gamma}$ was estimated.  The results appear in the boxplot in Fig \ref{pattern_output}a (blue bars).  The median is near 2m at most crossing angles, but drops to 1.3m in the counterflow condition.  A one-way ANOVA found that $\Bar{\lambda}$ depended significantly on crossing angle $\alpha$, $F(5,100) = 3.426$, $p = 0.0067$, $\eta^2 = 0.15$. A trend analysis revealed significant linear through $5^\text{th}$-order trends (all $p$’s $< 0.01$), indicating that the relationship was irregular, not monotonic (see \nameref{S3_Fig}). In the second analysis, we fit each group independently to estimate the stripe spacing for the left and right groups, $\Tilde{\lambda}_L$, $\Tilde{\lambda}_R$ (see Fig \ref{schem_3M}b), at the same instant. These results also appear in Fig \ref{pattern_output}(a) (brown and orange bars), and exhibit a similar drop in stripe width in the counterflow condition.  One-way ANOVAs found a significant effect of crossing angle on $\Tilde{\lambda}_L$,  $F(5, 100) = 3.817$, $p = 0.0033$, $\eta^2 = 0.16$, but not on $\Tilde{\lambda}_R$, $F(5, 100) = 1.781$, $p = 0.124$, $\eta^2 = 0.082$, likely due to the higher variability in the latter group. A trend analysis found no significant trends for either $\Tilde{\lambda}_L$ or $\Tilde{\lambda}_R$ (all $p$’s $> 0.08$, see \nameref{S3_Fig}).

To compare the overall spacing of the crowd with the separate spacing in each group, we computed the absolute difference between them on each trial, $|\Bar{\lambda}-\Tilde{\lambda}_L|$ and $|\Bar{\lambda}-\Tilde{\lambda}_R|$.  The median differences (Fig \ref{pattern_output}b) are generally  between 0.5m and 1m for acute and orthogonal crossing angles, but less than 0.2m for obtuse crossing angles. One-way ANOVAs confirmed that these differences significantly depended on crossing angle: for $|\Bar{\lambda}-\Tilde{\lambda}_L|$ , $F(5, 100) = 3.648$, $p = 0.0045$, $\eta^2 = 0.154$; for $|\Bar{\lambda}-\Tilde{\lambda}_R|$, $F(5, 100) = 5.796$, $p < 0.001$, $\eta^2 = 0.224$. Finally, we compared the stripe spacing in the left and right groups on each trial by computing the time-average of the difference $\langle|\Delta\Tilde{\lambda}|\rangle_t=\langle|\Tilde{\lambda}_L-\Tilde{\lambda}_R|\rangle_t$, during the interval $(T_i+T_f)/2$ to $T_f$ for each time frame of our data.  The results suggest that stripe spacing differed between groups by more than 0.6m at acute crossing angles, but by less than 0.4m at larger angles (Fig \ref{pattern_output}c).  A one-way ANOVA on $\langle|\Delta\Tilde{\lambda}|\rangle_t$ confirmed a significant dependence on crossing angle, $F(5, 100) = 4.26$, $p = 0.0015$, $\eta^2 = 0.175$.

\subsection*{Stripe width, number, and size based on the Edge-cutting algorithm}

\begin{figure}[h!]
\begin{adjustwidth}{-2.0in}{0in}
    \centering
    \includegraphics[width=18cm]{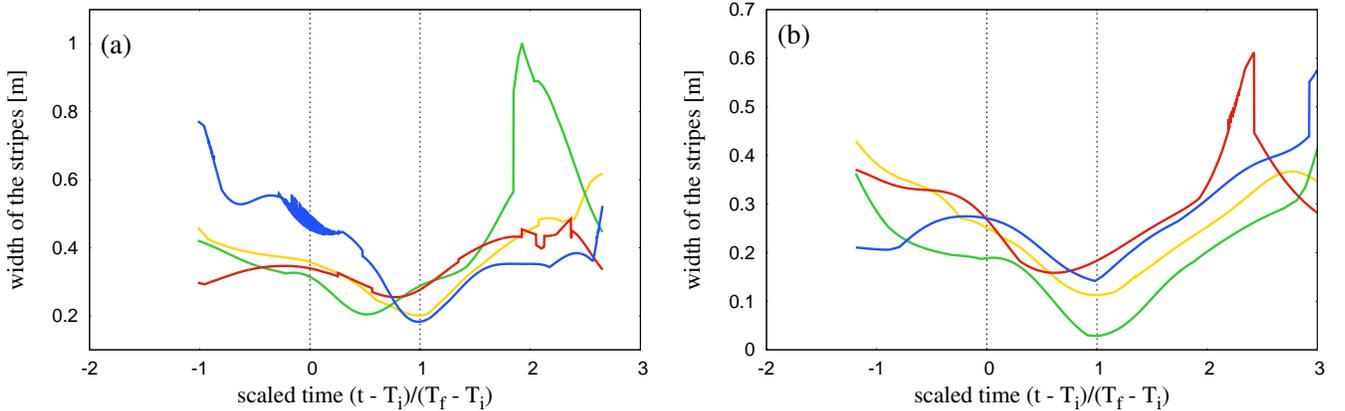}
    \vspace{0.5cm}
    \caption{\textbf{Width of stripes as a function of time.} Figure shows variation of stripe width as a function of time for all the stripes from a trial for two typical trials with (a) $\alpha=89.8\degree$ and (b) $\alpha=116.9\degree$. Time $t$ has been scaled as $t=\frac{t-T_i}{T_f-T_i}$. Thus, $t=T_i$ and $t=T_f$ correspond to the scaled values of 0 and 1 respectively, which are shown by vertical dashed lines in the figure. For almost all the cases, we see that the width of the stripe attains a global minimum within the interval 0 and 1, which represents the `squeezing' of stripes.}
    \label{squeeze}
\end{adjustwidth}    
\end{figure}

To characterize the dynamic behaviour of the stripes as the two groups crossed, we analyzed the variation in stripe width as a function of time. Precisely, we wanted to investigate the dynamic adjustments made by the pedestrians within a stripe to accommodate the incoming pedestrians from the other group. Once individual stripes were identified with the Edge-cutting algorithm, the width of each stripe was estimated by constructing a minimum bounding box for the stripe and taking its width dimension (see Materials and Methods). The dynamic variation in stripe width is plotted as a function of scaled time in Fig \ref{squeeze} for two different trials. Stripe width decreases at the onset of edge-cutting (time = 0) to a minimum before the last edge is cut (time = 1) and then increases again, as if the stripes are `squeezed' in space during the crossing interval. 

\begin{figure}[h!]
\begin{adjustwidth}{0.0in}{0in}
    \centering
    \includegraphics[width=9cm]{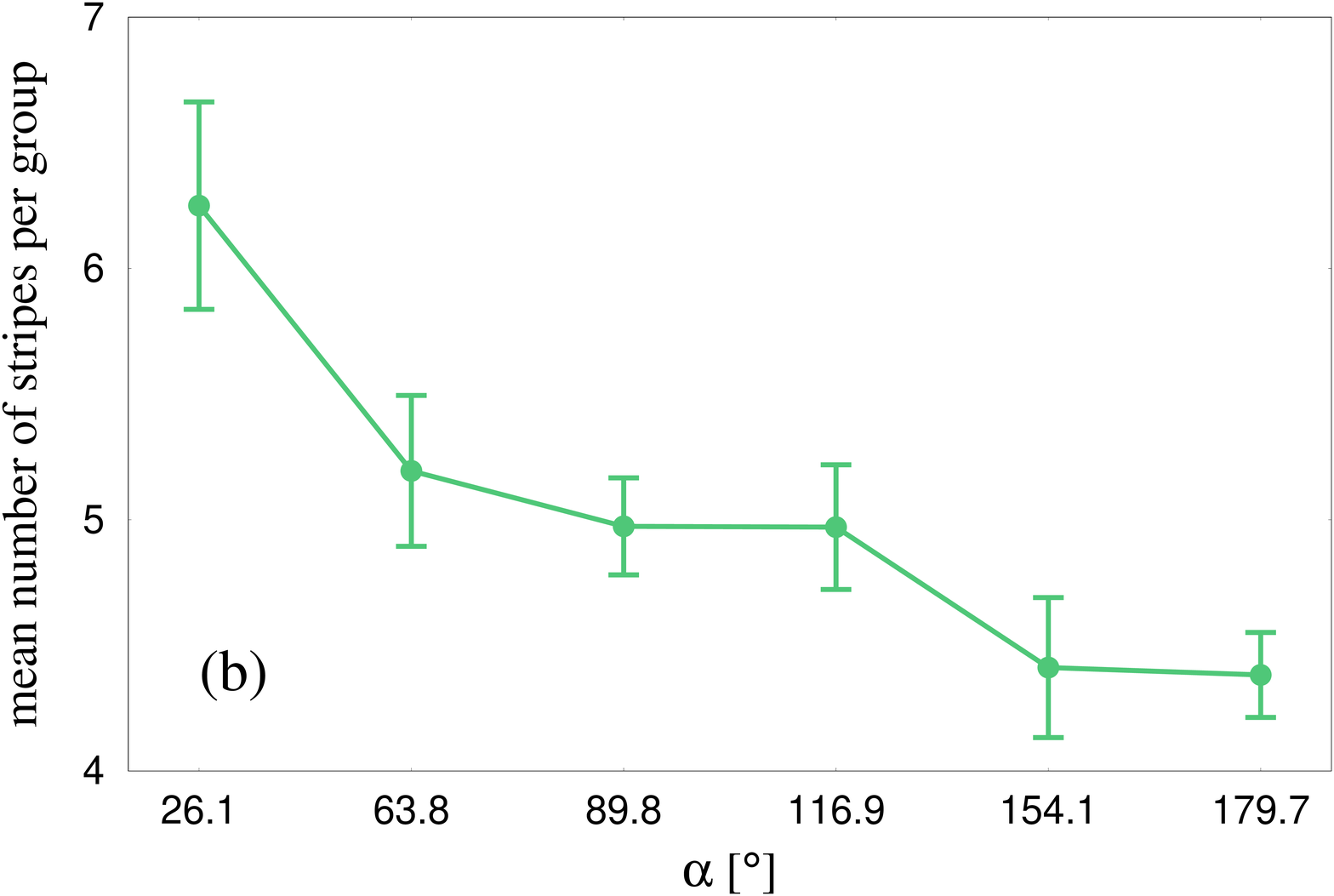}
    \vspace{0.5cm}
    \caption{\textbf{Mean number of stripes emerging from a group.} Figure shows the variation of this quantity with crossing angle $\alpha$. The mean was estimated over all the trials of our experiments. The number decreases with increasing $\alpha$. The error-bars indicate the corresponding standard errors of mean.}
    \label{mean_numb_str}
\end{adjustwidth}    
\end{figure}

The Edge-cutting algorithm also enables us to analyze the number of stripes that emerged during group crossing.  The mean number of stripes decreased monotonically as the crossing angle $\alpha$ increased, as represented in Fig \ref{mean_numb_str}.  This finding implies that stripe size (number of pedestrians per stripe) conversely increased with crossing angle, as suggested by the graph in \nameref{S5_Fig}.  Overall, the size of the identified stripes ranged from 1 to 15 participants in this study.

\section*{Discussion}

In this section we discuss the formation of striped patterns and their geometric properties as observed and estimated from our experimental data. This is followed by an evaluation of the two computational methods we used to derive our findings.

\subsection*{Did we observe stripe formation?}

\begin{figure}[h!]
\begin{adjustwidth}{-2.0in}{0in}
    \centering
    \includegraphics[width=9.5cm]{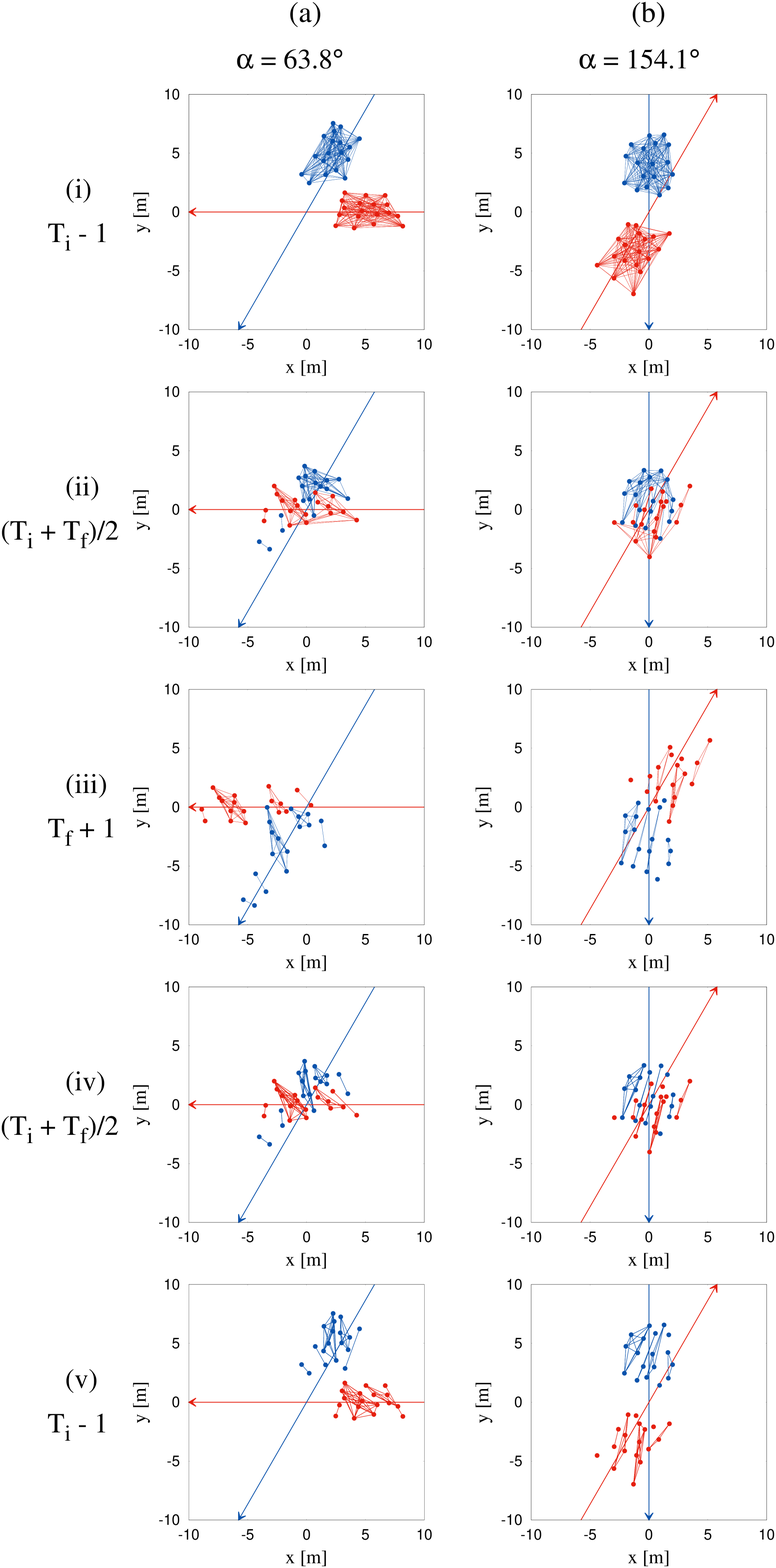}
    \caption{\textbf{Examples of the edge-cutting process.} Edge-cutting process for two trials with (a) $\alpha = 63.8\degree$
    and (b) $\alpha = 154.1\degree$.
    The blue and red arrows denote the directions of motion for the two groups of pedestrians shown by blue and red dots respectively. The instances shown in this figure goes forward in time from (i) to (iii) and backward in time from (iii) to (v). In (i) the instance shown is $T_i-1$, when all the edges within a group are intact. (ii) Shows the situation when the edges have started to cut and stripes are gradually being formed at $(T_i+T_f)/2$. (iii) Shows the situation at $T_f+1$ when all probable edge-cuts have taken place and the stripes have completely been formed. (iv) and (v) shows the instances as in (ii) and (i) respectively but with the visualisation of all the stripes that are completely formed only after $T_f$.}
    \label{ed60.150}
\end{adjustwidth}    
\end{figure}

Analyzing the formation of stripes was the main goal of this research.
In our experiments, we found that
stripe formation occurs even in small groups of pedestrians with fewer than 20 members, crossing in different directions without spatial constraints. This demonstrates that continuous flows in constrained channels are not necessary for self-organized pattern formation, which can be attributed to local interactions.
We should point out that,
there could have been a number of outcomes. For example, the two groups could have avoided without even penetrating each other, resulting in no formation of stripes. Large difference in velocities of the two groups could result in this scenario; thus the velocity of the two groups plays a crucial role in this context. Another possibility might be that crossing groups produced single isolated pedestrians, 
i.e., all the virtual connections between the pedestrians from one group were destroyed by pedestrians from the other group. This situation would also result in absence of stripe formation.
The two groups avoiding each other and the isolation of single pedestrians are the two extreme possibilities of outcomes from our experiments.
In reality we saw that the two groups indeed penetrate each other. The edge-cutting algorithm revealed the groups of around 20 participants, divided into 4 to 7 subgroups, as shown in Fig \ref{mean_numb_str}.
This confirms that not all pedestrians from a group end up being isolated as a consequence of crossing.
The identification of the participants belonging to a stripe was then used to calculate the orientation and width of the stripe. Several examples of the edge-cutting process are shown in Fig \ref{edge_ex} and \ref{ed60.150}.

\subsection*{Do the stripe properties depend on crossing angle?}

We were primarily interested in the effect of crossing angle on the orientation $\gamma$ of the stripes with respect to the bisector of the crossing angle. Based on previous observations and simulations, the expected value of $\gamma=90\degree$ should be invariant over different values of crossing angle $\alpha$. The results obtained from our experimental data using several methods of measurement are shown in Fig \ref{gamma_comp}a. The deviation of the median value of $\gamma$ remained less than $3\degree$ at all crossing angles, which is in good agreement with the bisector hypothesis.

Spatial separation $\lambda$ between two stripes from the same group was output from the pattern matching technique. We compared the values of $\lambda$ estimated using whole-crowd and separate-group analysis in Fig \ref{pattern_output}a as a function of $\alpha$.

Estimations of individual stripe properties based on the Edge-cutting algorithm revealed
that the mean number of stripes that emerges from a group decreases with $\alpha$, as shown in Fig \ref{mean_numb_str}. This implies that the mean size of a stripe should show an increase with $\alpha$. In \nameref{S5_Fig} we have shown the plot of mean size of a stripe as a function of the crossing angle $\alpha$. The mean stripe size indeed increases with $\alpha$. Thus the edge-cutting algorithm is very useful to establish the dependence of individual stripe properties on the crossing angle.

\subsection*{Comparison of assumptions and results for the whole-crowd and separate-group analyses using the pattern matching technique}

To perform pattern matching using a two dimensional sinusoid we make two different assumptions about the formation of stripes. In our analysis of finding $\Bar{\gamma}$ and $\Bar{\lambda}$, we assumed that the orientation of the two groups are parallel to each other and the stripes formed are periodic and equispaced. However we kept in mind that in reality this might not always be true. The orientation of the stripes for the two groups could be different and have different spacing. Thus, in a preliminary analysis we estimated stripe orientation
$\Tilde{\gamma}$ and their physical separation $\Tilde{\lambda}$ for the two groups separately as a function of time. In this approach we assumed that the stripes within one group are parallel to each other and equispaced. The time window which was selected for this analysis was from $T_i$ to $T_f$. The two timescales $T_i$ and $T_f$ were estimated from the Edge-cutting algorithm, and they approximately denote the beginning and end of interaction, between the two groups.

$\Tilde{\gamma}$ and $\Tilde{\lambda}$ for each groups
show some fluctuations near $T_i$ i.e. when the stripes have just started to form (see \nameref{S4_Fig}). The fluctuations reduce with time and the values of $\Tilde{\gamma}$ and $\Tilde{\lambda}$ approach a more steady value near $T_f$ i.e. when the stripes have been formed. Thus we calculate $\langle\Tilde{\gamma}_L\rangle_t$ and $\langle\Tilde{\gamma}_R\rangle_t$, i.e., the time averages of $\Tilde{\gamma}_L$ and $\Tilde{\gamma}_R$ (shown in Fig \ref{theta_avg}(a)) and the time-averaged difference
$\langle|\Delta\Tilde{\lambda}|\rangle_t=\langle|\Tilde{\lambda}_L-\Tilde{\lambda}_R|\rangle_t$ (shown in Fig \ref{pattern_output}c) from $(T_i+T_f)/2$ to $T_f$ i.e. when $\Tilde{\gamma}$ and $\Tilde{\lambda}$ remain approximately steady.

The differences of median values of $\langle\Tilde{\gamma}_L\rangle_t$ and $\langle\Tilde{\gamma}_R\rangle_t$ from the expected value of $90\degree$ are less than $2\degree$ for cases with an obtuse angle, and less than $8\degree$ for cases with an acute crossing angle with no statistical effect of crossing angle. This approximately justifies our earlier assumption about the orientation of stripes from two groups being parallel to each other. The time averaged difference in stripe spacing between the two groups  $\langle|\Delta\Tilde{\lambda}|\rangle_t$ also shows low median values - less than 0.8 m for all the crossing angles. 
This also justifies our assumption of parallel stripes from two groups when using the pattern matching technique.

We also make a comparative analysis of the differences in values of physical spacing of the stripes at the same time instant obtained by whole-crowd and separate-group analyses under the pattern matching technique. The results for $|\Bar{\lambda}-\Tilde{\lambda}_L|$ and $|\Bar{\lambda}-\Tilde{\lambda}_R|$ are shown in Fig \ref{pattern_output}b. The median values of the differences are less than 0.2 m for obtuse crossing angles, but are higher for acute crossing angles.

There was a possibility of `chevron' effect to create the differences in observed values of $\Tilde{\gamma}_L$ and $\Tilde{\gamma}_R$.
But due to the small size of our groups, the pedestrians might tend to move faster while leaving the crossing region - resulting in the absence of chevron effect.
The non-uniformity in the velocities of the agents both within the group and across the groups could also lead to deformation of stripes. One explanation could be the duration of time when the two groups keep interacting with each other. For lower values of $\alpha$ this duration is higher (see \nameref{S6_Fig}),
which results in deformation in the symmetric structure of the stripes. There could also be an effect of the size of the environment where the experiments were performed. Because of limitation of space used for the experiments, the two groups start interacting immediately after the commencement of trials for lower crossing angles. So there is a possibility that the agents participating for trials with acute crossing angle (e.g. $30\degree$) 
are still accelerating when reaching the crossing region, which clearly is not the case for trials with higher crossing angles.
To investigate this further one needs to perform an analysis with larger number of people in a bigger environment and eventually, with a flow of people - not just two groups crossing each other.

To analyse the statistical dependence of obtained results on the two methods
under pattern matching technique we performed ANOVAs for each of the crossing angles separately. For $\Bar{\gamma}$, $\Tilde{\gamma}_L$ and $\Tilde{\gamma}_R$ ANOVAs reveal no dependence of these quantities on the two methods 
for each of the crossing angles. For each of the cases the $p$-value is greater than 0.145. For $\Bar{\lambda}$, $\Tilde{\lambda}_L$ and $\Tilde{\lambda}_R$, the results were seen to be statistically independent of the two methods
except for the case $\alpha=89.8\degree$, as could also be seen from Fig \ref{pattern_output}a. Except $\alpha=89.8\degree$, the $p$-values are greater than 0.266. The results of the ANOVAs are shown in supplementary material (\nameref{S1_table} and \nameref{S2_table}).

\subsection*{Comparing results between pattern matching technique and the edge-cutting algorithm}

Using the edge-cutting algorithm we have conducted a per-stripe analysis, where properties of individual stripes were studied.
This helps us in a minimal way to explore the apparent asymmetry in the stripes from the two groups, which has been discussed earlier. The edge-cutting algorithm gives us the knowledge of stripes formed viz. the pedestrians belonging to a stripe. Using this output we compute the orientation and width of each of the stripes by implying rotating calipers algorithm (see Materials and Methods). The orientations $\gamma_L$ and $\gamma_R$ were computed as a time series. The time averaged orientation $\langle\gamma_L\rangle_t$ and $\langle\gamma_R\rangle_t$ were computed in the time interval $(T_i+T_f)/2$ to $T_f$. The boxplots over all the stripes and all the trials are shown in Fig \ref{theta_avg}b. The difference of the median values of these average quantities from the expected orientation (90$\degree$) are less than 5$\degree$ for obtuse crossing angles, but are a bit higher for acute crossing angles - a trend similar to previously discussed observations. The values of $\gamma_L$ and $\gamma_R$ computed at the same instant as when $\Bar{\gamma}$ were computed, are shown in Fig \ref{gamma_comp}a. In all the cases we observe that for obtuse crossing angles the stripe orientations obtained by the edge-cutting algorithm are not very different from that obtained by the pattern matching technique. However, for the acute crossing angles the differences are a bit higher - possible 
for reasons which have already been discussed. The width of the stripes as estimated from the per stripe analysis are not actually comparable to the physical separation of the stripes as computed by pattern matching technique. The stripes consisted of different numbers of people - this causes an irregularity while we attempt to compute their individual orientation. As a consequence, it would be inappropriate to compare these values with the outputs of the pattern matching technique, where the symmetry and periodicity of the stripes were assumed. However, we see that the median values of the average quantities $\langle\gamma_L\rangle_t$ and $\langle\gamma_R\rangle_t$ are not very far (approximately within 10$\degree$ for all crossing angles) from the expected value (90$\degree$), which was computed assuming the symmetry and regularity of the stripes. This shows consistency across the methods that we have used to study stripe properties.

\subsection*{Why did we use two different methods?}

The two computational methods that we present in this paper have
never been used before to study striped patterns in crossing flows, to the best of our knowledge. We used the two methods, viz. the edge-cutting algorithm and pattern matching technique, to study the formation and geometric properties of the stripes. The edge-cutting algorithm takes into consideration the entire trajectories of the pedestrians, whereas for the pattern matching technique the instantaneous positions of the pedestrians are sufficient.
Only the edge-cutting algorithm can identify a stripe and the pedestrians belonging to it. 
This yields a better definition of individual stripes and allows refined analysis of individual stripes, and is thus a spatially local method. Besides, this algorithm provides the full dynamics of individual stripes, and is the most appropriate to study dynamical effects such as the 'squeezing effect' of Fig \ref{squeeze} that we shall discuss shortly after.
When the stripes are very small (less than 3 participants) or are not sufficiently elongated (see Materials and Methods), their geometric properties are not well defined and we excluded them from our per stripe analysis.
On the other hand, the pattern matching technique uses a two-dimensional parametric sinusoid and is thus a spatially global method. This idea was inspired by Gabor functions, which have been used to model the spatial frequency response of the mammalian visual system \cite{marcelja}. We assume the existence of a periodic pattern of parallel stripes and then use this method to look for it; these assumptions have been borne out by the similarity of orientation and spacing when measured in the whole crowd and separately for the two groups.
For our small-scale
data the pattern matching technique is essential to study the orientation
of the stripes.

\subsection*{How efficient is the pattern matching technique?}

The pattern matching technique that we have used to find the orientation $\gamma$ of stripes and their spacial separation $\lambda$, was based on maximising $C$. $C$ is obtained by fitting a two-dimensional sinusoid $f$ on the coordinates of the pedestrians (see Materials and Methods). Therefore $C$ could be treated like a scoring function which indicates the quality of fitting. For the case when we assume that stripes from the two groups are parallel to each other and alternately equispaced (to find $\Bar{\gamma}$, $\Bar{\lambda}$), the maximising function is denoted by $\Bar{C}$ and when we fitted the two groups separately (to find $\Tilde{\gamma}$, $\Tilde{\lambda}$), this function was denoted by $\Tilde{C}$.

\subsubsection*{Importance of $C$ and $\lambda$}

\begin{figure}
\begin{adjustwidth}{-2.0in}{0in}
    \centering
    \includegraphics[width=18cm]{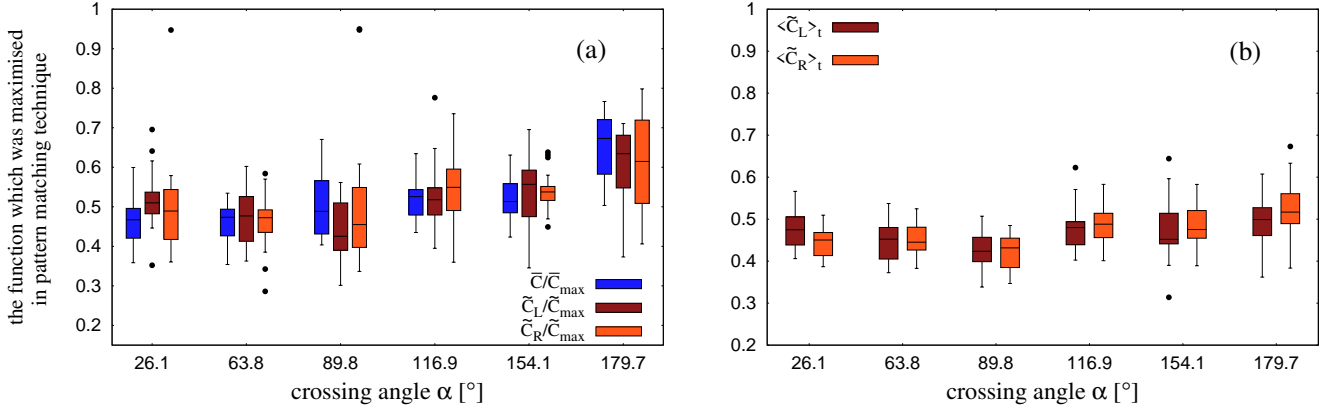}
    \vspace{0.2cm}
    \caption{\textbf{Boxplots for the maximising function of pattern matching technique.} (a) Boxplots for $\Bar{C}/\Bar{C}_{max}$ and $\Tilde{C}/\Tilde{C}_{max}$, where $\Bar{C}$ is the maximising function of pattern matching procedure of whole-crowd analysis and $\Tilde{C}$ is the same with separate group-analysis. $\Bar{C}_{max}$ and $\Tilde{C}_{max}$ are the maximum possible values of the maximising functions in these two cases, which are 2 and 1 respectively. (b) Boxplots for the time-averaged values of $\Tilde{C}_L$ and $\Tilde{C}_R$.}
    \label{C1C2}
\end{adjustwidth}    
\end{figure}

For best fittings one would get $\Bar{C}=\Bar{C}_{max}=2$ and $\Tilde{C}=\Tilde{C}_{max}=1$, and for the worst case (disordered input points) the sinusoidal function would not fit - it would either over-fit or under-fit the data points. Over-fitting or under-fitting could be identified by the obtained value of spatial separation $\lambda$ between the stripes. The obtained value of $\lambda$ was therefore very crucial to justify the pattern matching technique. From edge-cutting algorithm one could have an approximate idea of the spatial separation between two stripes for a trial (see Materials and Methods). $\lambda$ would be very low or very high compared to this approximate value in case of over-fitting and under-fitting respectively. The quality of the pattern matching technique is therefore estimated both in terms of $C$ and by the optimised value of $\lambda$. Fig \ref{edge_gabor} (right) shows the variation of $\Bar{C}$ as a function of $\Bar{\lambda}$ for two typical trials. In Fig \ref{C1C2} we have shown boxplots for the values of $C$ for each of the crossing angles, as obtained by whole-crowd and separate-group analysis under the pattern matching technique. Higher values of $C$ indeed signify a better fitting. From Fig \ref{C1C2} we see that the median values of both $\Bar{C}$ and $\Tilde{C}$ increase with $\alpha$. One-way ANOVAs on $C$ found a significant effect of crossing angle on $\Bar{C}$, $F(5, 100) = 17.53$, $p < 0.001$, $\eta^2 = 0.467$, on $\Tilde{C}_L$, $F(5, 100) = 7.955$, $p < 0.001$, $\eta^2 = 0.285$, and on $\Tilde{C}_R$, $F(5, 100) = 3.665$, $p = 0.0043$, $\eta^2 = 0.155$.

\subsubsection*{Estimating residual error of the fitting}

\begin{figure}
\begin{adjustwidth}{0.0in}{0in}
    \centering
    \includegraphics[width=10cm]{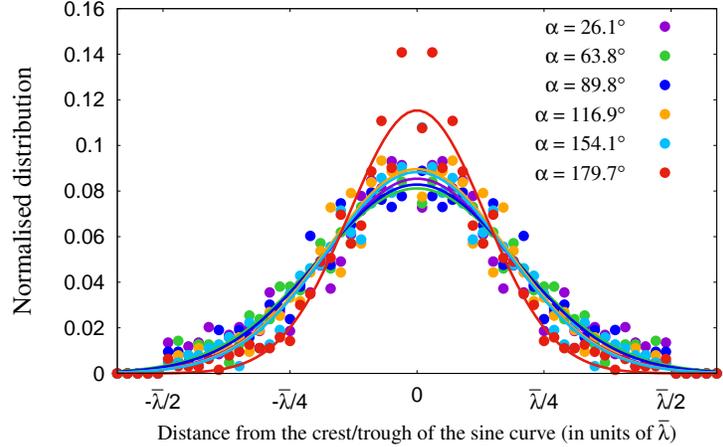}
    \vspace{0.3cm}
    \caption{\textbf{Accuracy of the pattern matching procedure.} Figure shows Normalised distributions for the distances of pedestrian positions from the crest or troughs of the fitted 2D sinusoid i.e. the residual errors. The distributions show a Gaussian peak at the origin for each $\alpha$. The data were fitted according to a Gaussian curve and the fitted curves are shown by solid lines. }
    \label{pdfres}
\end{adjustwidth}    
\end{figure}

To study the accuracy of the pattern matching procedure to find $\Bar{\gamma}$ and $\Bar{\lambda}$, we calculate the residual errors. Ideally one would expect all the data points to lie within the distance $-\Bar{\lambda}/4$ to $\Bar{\lambda}/4$, where $\Bar{\lambda}$ is the wavelength of the fitted sine curved $f$. We calculated the residual error of pattern matching technique as the distance of the data points from the crest or trough of the fitted sine function $f$. The results are shown in Fig \ref{pdfres}. The normalised distribution of this distance shows a Gaussian peak at the origin. We fit the data for each of the cases using the functional form of Gaussian distribution. From the fittings, we estimate the standard deviations. For $\alpha=179.9\degree$ the standard deviation of the fitted curve was $0.134\Bar{\lambda}$ and for the remaining crossing angles this value is $0.184\Bar{\lambda}$ on average. From the data of residual error, we found that for $\alpha=179.6\degree$, 92.4$\%$ of the data points are accumulated between the distances $-\Bar{\lambda}/4$ and $-\Bar{\lambda}/4$, and for the remaining crossing angles, on an average 85.3$\%$ of the data is within this range. This surely establishes the efficiency of pattern fitting to a great extent. Besides, this also underlines a difference in stability between lanes and stripes (discussed later).

\subsubsection*{Periodicity of the two groups}

The periodic arrangement of stripes that are seen to form in our experiments have been a point of concern for the pattern matching technique. An important aspect of our pattern fitting procedure is to choose the instant of time for which the position of pedestrians are considered and fitted. For higher values of $\alpha$ this instant is usually when all the edges have been cut and all possible clusters have been formed, which is $T_f$ - an output from the Edge-cutting algorithm. However, for lower values of $\alpha$ the periodicity of the two groups of pedestrians appears to be destroyed at $T_f$. For such trials, a suitable time is chosen which is less than $T_f$ but higher than $(T_i+T_f)/2$. The principal motivation of choosing such a time-scale is to protect the periodicity of the two groups and to ensure the formation of stripes up to such an extent that the position of pedestrians could be considered for sinusoidal fitting (see Materials and Methods). In Fig \ref{edge_gabor} we have shown pedestrian positions from the two groups for two typical trials and the fitted sine curves $f$, demonstrating the periodicity of the stripes.

\subsection*{Comparison of lanes and stripes}

The striped patterns that are seen for counter flows (i.e. $\alpha=179.6\degree$ for our case) are known in the literature as lanes. Our results confirm that lanes (which are parallel to the direction of motion) are more stable than stripes (which in general are not aligned with the direction of motion). From Fig \ref{pattern_output}, we see that all $\lambda$ measurements coincide better for counter flows and for this structure, the pedestrians accept a lower distance between the lanes (minimum of $\lambda$). Typically the distance between neighboring stripes is of the order of 0.8 to 1.1 m when the motion of pedestrians is not parallel to the direction of the stripes. While for lanes, the distance between the centers of lanes is rather 0.6 m - which, given the width of pedestrians, seems close to the minimum possible value if pedestrians intend to avoid collision. Higher stability of lanes compared to the stripes was also established when we estimate the residual error of sinusoidal fitting. We found that, $7\%$ more of data points lie between the expected range for counter flows, than compared to the other crossing angles, as could also be seen from Fig \ref{pdfres}.

\subsection*{Squeezing behaviour of the stripes: future investigations}

The macroscopic dynamics of the stripes, accessible thanks to the edge-cutting algorithm,
also show intriguing behaviour. When the two groups cross each other, the stripes that are formed get `squeezed' in order to accommodate in space the incoming group. Microscopically, each of the pedestrians within a stripe adjust their motion when they encounter a pedestrian from the opposite group. In Fig \ref{squeeze}
we have shown width of all the stripes from two typical trials as a function of a scaled time for each crossing angle $\alpha$. The time is scaled in such a way that the scaled value of 0 and 1 correspond to $T_i$ and $T_f$ of the trial. It is observed from Fig \ref{squeeze} that between the interval 0 and 1 i.e. the beginning and end of interactions, the width of the stripes decreases, attains a global minimum and then increases again. This indicates some interesting underlying microscopic behaviour of the agents, which results in the squeezing behaviour as a macroscopic property of the stripes. In our subsequent research we would be interested to determine the underlying mechanism responsible for this behaviour. It would also be appealing to find out whether a following behaviour is present among the pedestrians leading to the formation of stripes, which we plan to work in our next research.

\subsection*{Conclusion}

We conducted experimental trials for crossing flows of pedestrians without any spatial constraints of motion. In spite of having small number of participants we observed the formation of emerging striped patterns for each value of the crossing angle. Edge-cutting algorithm was implemented to detect
the formation of stripes. 
Striped patterns for counter flows i.e lanes are seen to be more stable than those for other crossing angles.
We have used a pattern matching technique and the edge-cutting algorithm to study a few properties of the stripes formed and compare them with each other and with hypothesized effects. The observed values for the orientation of stripes from edge-cutting algorithm are in good 
agreement with the expected result which justifies that our assumption about the regularity and symmetry of the striped patterns are reasonable enough.
The maximised values of $\Bar{C}$ as obtained by us signify the regularity of the striped patterns from the two groups. While performing numerical simulations to model the scenario of crossing flows, the quantity $\Bar{C}$ would act as a parameter to evaluate the effectiveness of the simulation technique in reproducing the observed behaviour.
We not only confirmed that stripe orientation is predicted by the bisector hypothesis at all crossing angles, 
but we also discovered several unexpected effects. First we showed that the average number of stripes within a group decreases with the crossing angle alpha. Second, we found that the spacing, number, and size of stripes depended significantly on crossing angle. Third, we observed a squeezing effect visible in the time evolution of the stripes. The macroscopic dynamics of the stripes motivates us to study the microscopic behaviour of the individual pedestrians as our next investigation.

\section*{Materials and methods}
\subsection*{Experimental details}

\begin{figure}
\begin{adjustwidth}{0.0in}{0in}
    \centering
    \includegraphics[width=7cm]{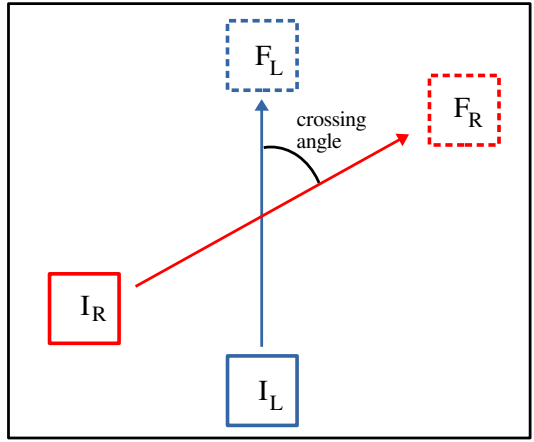}
    \vspace{0.3cm}
    \caption{\textbf{Schematic representation of the experimental set-up.} Figure shows the experimental set-up that we have constructed to study crossing flows of two groups of people. The squares ($I_L$ and $I_R$) with solid curves indicate the initial ($I$) positions and the squares ($F_L$ and $F_R$) with dashed curves represent the final ($F$) positions of the two groups. The suffixes $L$ and $R$ indicate the groups which are on the left ($L$) or right $R$ side of the bisector of the
    crossing angle.}
    \label{setup}
\end{adjustwidth}    
\end{figure}

The participants of the experiments were divided into two groups (with similar spatial densities). They were instructed to move along a direction which was announced before the commencement of each trial, such that the two groups cross each other at a particular angle. For 7 different expected values of crossing angles, viz. [$0\degree$, $30\degree$, $60\degree$, $90\degree$, $120\degree$, $150\degree$ and $180\degree$], we performed approximately 17 trials at each angle, a total of 116 trials (See Table \ref{summary}). During each trial the head trajectory of each pedestrian was recorded as a time series. Each trial lasted about 15-25 seconds.
The experiment was performed in a rectangular hall (20m x 30m) with a tracking area of 15m x 20m.
The positions of the pedestrians were recorded at 120 Hz using VICON - an infrared camera system.
The pedestrians were equipped with head-mounted reflective markers detectable by the VICON motion capture system.
The center of the tracking was considered as the origin of a two-dimensional Cartesian coordinate system, which was used as a reference to represent the position of the pedestrians at every time step. Table \ref{summary} summarizes the various details of the experiments, i.e. the number of pedestrians and number of trials for each value of the crossing angle. In Fig \ref{setup} we have schematically shown our experimental set-up.

For our experiments, we searched for participants on campus of University of Rennes, France. The participants had no visual or locomotive impairments. The experiments were performed over 2 days. On the first day, we could gather 36 participants and on the next day the number was 38. Before each trial, the group segregation was done randomly. The participants were unaware of the actual motivation of the experiments.

The data obtained from the experiments were low-pass filtered to reduce oscillations due to the gait movement of the walking pedestrians. We used a forward-backward 4-th order butterworth filter to reduce these unwanted oscillations. The traces of pedestrians shown in Fig \ref{traj} are plotted using the filtered trajectories. For all the analysis presented in this paper, we have used the filtered data.

\begin{table}
\begin{adjustwidth}{0.0in}{0in}
\centering
\caption{\textbf{Summary of the experimental details}}
\begin{tabular}{c|c|c|c}
\cline{1-4}
{Expected} & {Observed} & {No. of} & {No. of}\\
{crossing} & {crossing} & {pedestrians} & {trials}\\
{angle} & {angle $\alpha$} & {} & {}\\

\cline{1-4}

{\multirow{2}{*}{$0\degree$}} & {\multirow{2}{*}{}} &
{38} & {6} \\
{} & {} & {36} & {4}\\
\cline{1-4}

{\multirow{2}{*}{$30\degree$}} & {\multirow{2}{*}{$26.1\degree$}} & {38} & {10}\\
{} & {} & {36} & {8}\\
\cline{1-4}

{\multirow{2}{*}{$60\degree$}} & {\multirow{2}{*}{$63.8\degree$}} & {38} & {10}\\
{} & {} & {36} & {8}\\
\cline{1-4}

{\multirow{2}{*}{$90\degree$}} & {\multirow{2}{*}{$89.8\degree$}} & {38} & {11}\\
{} & {} & {36} & {8}\\
\cline{1-4}

{\multirow{2}{*}{$120\degree$}} & {\multirow{2}{*}{$116.9\degree$}} & {38} & {10}\\
{} & {} & {36} & {7}\\
\cline{1-4}

{\multirow{2}{*}{$150\degree$}} & {\multirow{2}{*}{$154.1\degree$}} & {38} & {9}\\
{} & {} & {36} & {8}\\
\cline{1-4}

{\multirow{2}{*}{$180\degree$}} & {\multirow{2}{*}{$179.7\degree$}} & {38} & {10}\\
{} & {} & {36} & {7}\\
\cline{1-4}
\end{tabular}
\begin{flushleft}
\vspace{0.5cm}
{Table 
indicates number of pedestrians and number of trials categorized according to the value of the crossing angle. The observed values of the crossing angle which are mentioned here are basically the median values over all the trials.}
\end{flushleft}
\label{summary}
\end{adjustwidth}
\end{table}

\subsection*{Ethics Statement}

The ethical approval for using live participants was obtained from \textit{The Operational Committee for the Evaluation of Legal and Ethical Risks} (COERLE - n$\degree$ 2016-008). The document could be found at \nameref{S1_Doc}. Written consents were taken from the participants who volunteered for the experimental trials.

\subsection*{Observed values of crossing angle $\alpha$}

During the experiments the participants followed visual references for their movement, instead of a secluded corridor. As a consequence the actual direction of motion of the groups and hence the actual value of the crossing angle is a bit different from what it was expected i.e. the 
expected values. Therefore we calculate the observed values of crossing angle $\alpha$ (see \nameref{S1_Fig}), and show all of our findings in terms of them.
To calculate the observed values of the crossing angle $\alpha$ we consider the two barycenters of the initial and final positions of all the participants in a group for a trial. The line connecting these two points gives the actual direction of motion of a group, from which we evaluate the observed crossing angle $\alpha$.  We then compute medians over all the trials and use these median values in all of our analysis. In Table \ref{summary} we mention the median values of the observed crossing angle $\alpha$
We also expect the individual pedestrians to make some personal adjustments in their trajectories to reach their goal. Thus we measure how much a pedestrian actually deviates from his/her originally assigned trajectory. The normalised distributions of this measurement show Gaussian behaviour (see \nameref{S2_Fig}). The mean value of the angular deviations in each case is less than $2\degree$.

\subsection*{Edge-cutting Algorithm: Detection of the stripes}

\begin{figure}[h!]
\begin{adjustwidth}{0.0in}{0in}
    \centering
    \includegraphics[width=7cm]{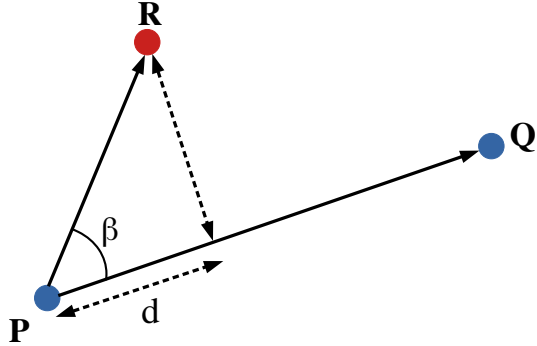}
    \vspace{0.3cm}
    \caption{\textbf{Schematic representation of the edge-cutting algorithm.} The `edge' between the pedestrians $P$ and $Q$ belonging to the same group is cut by the pedestrian $R$ belonging to the other group.}
    \label{schem_edgecut}
\end{adjustwidth}    
\end{figure}

In the beginning of the trial, at time $t=0$, we assume that each group of pedestrians forms a complete graph with clustering coefficient = 1 i.e. all the individuals are connected to each other within the group by an `edge'. The basis of such an assumption is the correlated movements of pedestrians in a group \cite{henderson}. With the progression of time, when the two groups meet and cross each other, the edge between two pedestrians from one group may be cut by a pedestrian from the other group. This situation is detected by the edge-cutting algorithm. Once all the probable edge-cuttings are over, each group is left with more than one cluster having a complete graph. The size of these clusters are $\geq 1$. Fig \ref{schem_edgecut} schematically represents the scenario of the edge connecting the individuals $P$ and $Q$ from the same group being cut by the individual $R$ from the other group. The conditions for the edge being destructed are three-fold and as follows:\\
(i) $\overrightarrow{PQ}.\overrightarrow{PR} > 0$\\
(ii) $d = \overrightarrow{PQ}.\overrightarrow{PR} < |{\overrightarrow{PQ}}|^2$\\
(iii) $\beta(t)\times\beta(t-1) < 0$,\\
where the angle $\beta$ is defined in Fig \ref{schem_edgecut}.
Simultaneous satisfaction of these three conditions detects the edge-cutting. Conditions (i) and (ii) ensure that the pedestrian $R$ is able to cut the edge between $P$ and $Q$. When these two conditions are satisfied, the angle $\beta$ between $\overrightarrow{PQ}$ and $\overrightarrow{PR}$ is measured as a time series, and if it changes sign we confirm that the edge is destructed. Condition (iii) allows us to detect the time of edge-cutting as well. For a trial, the instant when the first (initial) edge-cut takes place is denoted by $T_i$ and the instant of the last (final) edge-cut is denoted by $T_f$. The two timescales $T_i$ and $T_f$ are outputs from the Edge-cutting algorithm, and they have been used extensively in the analysis of stripe orientations.

\subsection*{Pattern matching: Fitting 2D parametric sinusoidal curves}

\begin{figure}[h!]
\begin{adjustwidth}{-0.0in}{0in}
    \centering
    \includegraphics[width=12cm]{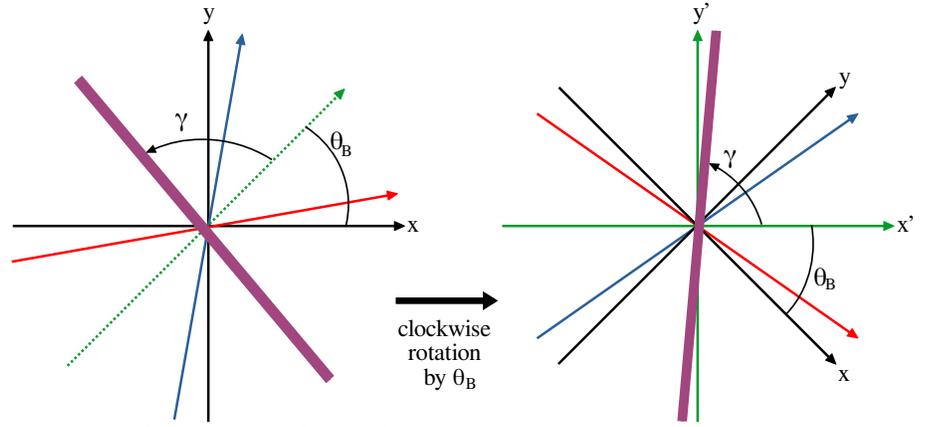}
    \caption{\textbf{Transformation of coordinates.} This diagram schematically represents the transformation given to the experimentally obtained data such that the transformed $x$-axis, i.e., $x'$ is along the bisector of the two group direction vectors. The arrows in blue and red indicate the two group direction vectors and the dotted green arrow indicates the bisector. $\theta_B$ is the angle between the bisector and the original $x$-axis. A clockwise rotation by an angle $\theta_B$ in this case makes the bisector as the new $x$-axis. The transformed axes $x'$ and $y'$ are shown by green arrows. The bold line in purple represents a stripe, which makes an angle $\gamma$ (measured anti-clockwise) with the bisector of group direction or the $x'$-axis. The purpose of the pattern matching technique is to find out the orientation of stripes $\gamma$.}
    \label{transform}
\end{adjustwidth}
\end{figure}

To estimate the orientation of the parallel stripes and their physical separation we implied a pattern matching technique. We use a two dimensional parametric sinusoidal function $f$ for this method and fit this function on the pedestrian positions. The goal of the pattern matching technique was to (i) estimate the angle $\gamma$ between the stripes and the bisector of the crossing angle and (ii) to estimate the physical separation $\lambda$ between the stripes from the same group. In all the cases discussed in this paper, the orientation $\gamma$ of the stripes were measured counterclockwise from the bisector of the crossing angle $\alpha$. The data obtained by experiments were given a transformation such that $x'$, the new $x$-axis coincided with the bisector of the crossing angle, this is illustrated
in Fig \ref{transform}. This transformation was applied so that the orientation $\gamma$ could be directly evaluated from pattern matching. The function $f$ was fitted on the transformed coordinates of the pedestrians $(x',y')$. The fitting was achieved by maximising a function $C$, which is the mean of sum of values of $f$ as fitted on the pedestrian positions.

For the pattern matching procedure when we assume that the stripes from the two groups are parallel to each other and 
are equispaced, we 
denote quantities with a bar as a way of representation. In this case the sinusoidal function $f$ was given by the form
\begin{equation}
    f(x',y';\Bar{\gamma},\Bar{\lambda},\psi)=\sin\Big(\frac{2\pi X}{\Bar{\lambda}}+\psi\Big)
    \label{sinf}
\end{equation}
where, $X=x'\sin\Bar{\gamma}-y'\cos\Bar{\gamma}$, $\Bar{\lambda}$ is the wavelength of the sine function and $\psi$ denotes the phase offset. To find the orientation $\Bar{\gamma}$ of the parallel stripes, we fit $f$ to the position $(x',y')$ of the pedestrians. The fitting was performed using an optimisation strategy, where we maximise $\Bar{C}$, which is defined as
\begin{equation}
\Bar{C} = \Big(\sum_{\text{ group 1}}f(x',y')+\sum_{\text{ group 2}}-f(x',y')\Big)/\Bar{N},
\label{Cfunc}
\end{equation}
where $\Bar{N}$ is the total number of pedestrians. The first summation sign denotes the sum over the position of pedestrians from one group and the second summation for the position of pedestrians from the other group. The maximum possible value of $\Bar{C}$ is 2, which occurs for the ideal case when the position of pedestrians from the two groups could be fitted exactly on the crests and troughs of the sinusoidal curve respectively. Maximisation of $\Bar{C}$ by fitting $f$ on pedestrian positions gives us the orientation $\Bar{\gamma}$ of the stripes and spatial separation $\Bar{\lambda}$ between two stripes from the same group. Evaluation of $\Bar{\gamma}$ and $\Bar{\lambda}$ is done under the assumption that the stripes from the two groups are parallel to each other and are equispaced. Pictorial representation of $\Bar{\gamma}$ and $\Bar{\lambda}$ is shown in Fig \ref{schem_3M}a.

For a randomly oriented set of points, the pattern matching technique would over-fit or under-fit the data - that could be detected by the obtained value of $\Bar{\lambda}$. In our case, before performing the pattern matching procedure, we imply the edge-cutting algorithm to obtain the number of stripes that are formed in a trial. Combining this knowledge with the width of the crossing region, we get an approximate estimate for the value of spatial separation $\Bar{\lambda}$. This helps us to identify any over-fitting or under-fitting. For the trial with $\alpha=89.9\degree$ shown in Fig \ref{edge_ex} and Fig \ref{edge_gabor}, the width of the crossing region is 8.06 m. From the edge-cutting algorithm we get that the two groups of this trial gets divided into a total of 9 subgroups. Therefore, approximate estimate for the wavelength of fitted sinusoid is $2\times\frac{8.06}{9}=1.79$ m. From the pattern fitting we obtained $\Bar{\lambda}=1.865$ m, which is quite close to the approximate estimate. The fitted sine functions $f$ for two typical trials and variation of $\Bar{C}$ as a function of $\Bar{\gamma}$ and $\Bar{\lambda}$ for the same trials are depicted in Fig \ref{edge_gabor}.

To find the orientation of stripes for the two groups in a trial separately, we used the same fitting function $f$ as in Eq (\ref{sinf}) and the function which was maximised to obtain the fitting in this case was 
\begin{equation}
    \Tilde{C} = \sum f(x',y') / \Tilde{N},
\end{equation}
where the summation was performed over the position of pedestrians from one group at a time. $\Tilde{N}$ is the number of pedestrians in the group. The maximum possible value of $\Tilde{C}$ is 1, which in this case occurs for the ideal situation when the pedestrian positions from the group under consideration fall exactly on the crests of the sine curve represented by $f$. This analysis was performed as a time sequence between $(T_i+T_f)/2$ and $T_f$. Maximisation of $\Tilde{C}$ by fitting $f$ on the pedestrian positions gives us the orientation $\Tilde{\gamma}$ of the stripes and the spatial separation $\Tilde{\lambda}$ between the stripes from the same group. This computation was done under the assumption that the stripes from the same group are parallel to each other and have equal spacing between them. 

While fitting the parallel stripes from the two groups separately we differentiate them by using the notations $\Tilde{\gamma}_L$ and $\Tilde{\gamma}_R$. $\Tilde{\gamma}_L$ denotes the orientation of the stripes whose group direction vector lies to the left (L) of the direction of bisector and similarly, $\Tilde{\gamma}_R$ for the one whose group direction vector lies to the right (R) of the direction of bisector. Similarly, we use the notations $\Tilde{\lambda}_L$ and $\Tilde{\lambda}_R$ to denote the spatial separation between the stripes from the same group, according to the orientation of its group direction vector with respect to the bisector. Pictorial demonstration of $\Tilde{\gamma}_L$, $\Tilde{\gamma}_R$, $\Tilde{\lambda}_L$ and $\Tilde{\lambda}_R$ are shown in Fig \ref{schem_3M}b. Following the same convention, the functions that were maximised to obtain ($\Tilde{\gamma}_L$, $\Tilde{\lambda}_L$) and ($\Tilde{\gamma}_R$, $\Tilde{\lambda}_R$) were denoted as $\Tilde{C}_L$ and $\Tilde{C}_R$ respectively.

For the segregation of the groups according to whether they lie to the left or right of the bisector of the crossing angle, it is therefore important to determine the direction of the bisector. This direction is estimated using the two group direction vectors. But for the case of crossing angle $180\degree$, determining the direction of the bisector is not possible. However we realise that the experimentally observed value of the crossing angle $\alpha$ is never exactly equal to 180$\degree$. Thus estimating the direction of bisector for these cases is also pretty straight-forward.

\subsection*{Finding individual stripe width and orientation}

\begin{figure}[h!]
\begin{adjustwidth}{0.0in}{0in}
    \centering
    \includegraphics[width=12cm]{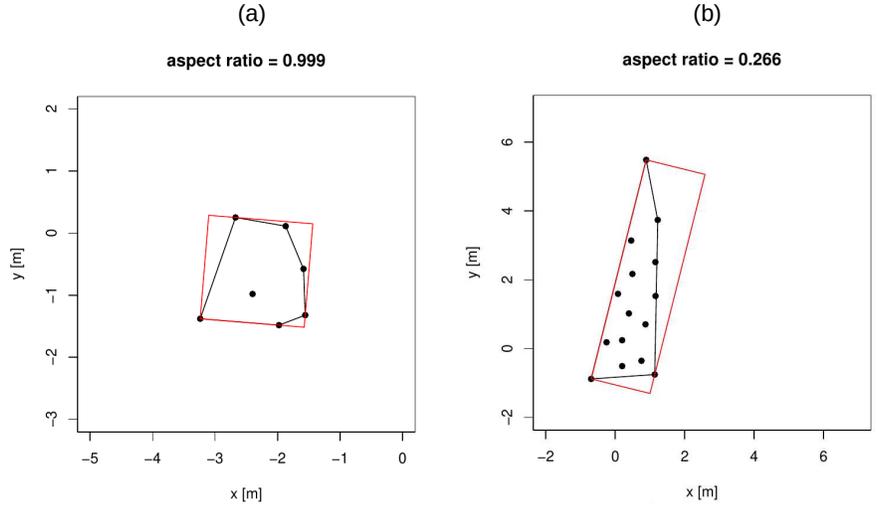}
    \caption{\textbf{Constructing the minimum bounding box of a stripe.} Two typical stripes are shown with aspect ratios (a) 0.999 and (b) 0.266. The red boxes denote the minimum bounding boxes and the polygons shown by black lines are the convex hulls of the points in the stripes. The stripe with the aspect ratio closer to 1 (a) is not suitable for the estimation of orientation. The stripe with the lower aspect ratio (b) is inclined at an angle of $76.01\degree$ with respect to the $x$-axis.}
    \label{asp}
\end{adjustwidth}    
\end{figure}

From the Edge-cutting algorithm we could successfully identify the stripes that are formed. In our attempt to find the individual stripe orientations at each instant we construct the minimum bounding box of the stripes using Rotating Calipers algorithm \cite{shamos_rca,toussaint}. The orientation of the stripe was calculated along the length of the box. The width of the rectangular box gives an estimate of the width of each of the stripes. The aspect ratio of the minimum bounding box for a stripe, calculated as the ratio of its width to length, gives an idea of the suitability of that stripe to be considered for the estimation of orientation. The value of aspect ratio closer to 1 denotes a uniformly shaped stripe. Whereas, lower value of aspect ratio indicates a sufficiently elongated stripe suitable for finding the orientation. In Figure \ref{asp}, we show two typical stripes with their respective minimum bounding boxes calculated using the Rotating Calipers algorithm. We applied a cut-off on aspect ratios of the stripes and considered only those stripes which had an aspect ratio less than 0.5. The time window which was selected for this calculation was from $(T_i+T_f)/2$ to $T_f$. The orientation of individual stripes were also estimated as the angle between the stripes and the bisector of the group direction vectors, as depicted in Fig \ref{schem_3M}c. The angle is measured counterclockwise from the bisector. We use the notations $\gamma_L$ and $\gamma_R$ to differentiate the orientation of stripes whose group direction vector lie to the left and right of the bisector respectively.

\section*{Supporting information}

\paragraph*{S1 Table.}
\label{S1_table}
{\bf Results of ANOVAs for $\Bar{\gamma}$, $\Tilde{\gamma}_L$ and $\Tilde{\gamma}_R$ for each $\alpha$.} Table summarizes the results of the ANOVA tests that were performed for each $\alpha$ with $\Bar{\gamma}$, $\Tilde{\gamma}_L$ and $\Tilde{\gamma}_R$ to check their statistical dependencies on whole-crowd and separate-group analyses under the pattern matching technique.

\paragraph*{S2 Table.}
\label{S2_table}
{\bf Results of ANOVAs for $\Bar{\lambda}$, $\Tilde{\lambda}_L$ and $\Tilde{\lambda}_R$ for each $\alpha$.} Table summarizes the results of the ANOVA tests that were performed for each $\alpha$ with $\Bar{\lambda}$, $\Tilde{\lambda}_L$ and $\Tilde{\lambda}_R$ to check their statistical dependencies on whole-crowd and separate-group analyses under the pattern matching technique.

\paragraph*{S1 Fig.}
\label{S1_Fig}
{\bf Observed values of the crossing angle $\alpha$.} Figure shows boxplots for the observed values of the crossing angle $\alpha$. The dashed lines denote the corresponding values of the expected crossing angle.

\paragraph*{S2 Fig.}
\label{S2_Fig}
{\bf Normalised distributions of $\delta$ for each $\alpha$.} The quantity $\delta$ is a measure of the deviations that a pedestrian makes with his/her originally instructed direction of motion as indicated by the visual references during the trials. We calculate $\delta$ at every instant of a pedestrian's trajectory as the angle between the trajectory of that pedestrian and his/her expected direction of motion. Anti-clockwise (clockwise) deviations are considered as positive (negative). $\delta$ is estimated for all the pedestrians at all the positions along their trajectories. Normalised distributions of $\delta$ is shown in this figure. The data for each $\alpha$ were fitted according to the curve $f(x)=a\exp{[-b(x-c)^2]}$. The quantity $c$ (in unit of $\degree$) gives the mean of each distribution. The solid lines indicate the best fitting curves. For each $\alpha$ we get $|c|< 2\degree$.

\paragraph*{S3 Fig.}
\label{S3_Fig}
{\bf Mean values of $\Bar{\lambda}$, $\Tilde{\lambda}_L$ and $\Tilde{\lambda}_R$ as a function of $\alpha$.} Figure shows the mean values of the observed quantities $\Bar{\lambda}$, $\Tilde{\lambda}_L$ and $\Tilde{\lambda}_R$ as a function of the crossing angle $\alpha$. The error-bars indicate the corresponding standard errors of mean. Trend analyses show that dependence of $\Bar{\lambda}$ on $\alpha$ is irregular, not monotonic; whereas $\Tilde{\lambda}_L$ and $\Tilde{\lambda}_R$ show no significant trends.

\paragraph*{S4 Fig.}
\label{S4_Fig}
{\bf Time sequence of $\Tilde{\gamma}$ and $\Tilde{\lambda}$ estimated separately for the two groups in a trial.} The plots are shown for two typical trials with (a) $\alpha=63.8\degree$ and (b) $\alpha=154.1\degree$. The time window for which the data is shown is between the respective values of $T_i$ and $T_f$ for both the trials. $T_i$ and $T_f$ are evaluated the Edge-cutting algorithm.
For $\Tilde{\gamma}$ plots in the left panel, the dashed lines indicate 90$\degree$, the expected value of stripe orientation and for $\Tilde{\lambda}$ plots in the right panel, the dashed lines indicate the value of $\Bar{\lambda}$ as estimated from the pattern matching technique by considering the two groups together.

\paragraph*{S5 Fig.}
\label{S5_Fig}
{\bf Mean size of stripes as a function of $\alpha$.} The size of a stripes is defined as the number of pedestrians belonging to that stripe. The mean size of a stripe increases with the increase of crossing angle.
The error-bars indicate the corresponding standard errors of mean.

\paragraph*{S6 Fig.}
\label{S6_Fig}
{\bf Mean crossing time as a function of $\alpha$.} Crossing time for each trial is defined as $T_f-T_i$, where $T_i$ and $T_f$ are estimated from the edge-cutting algorithm. $T_i$ and $T_f$ denotes the time when the first and last edge-cut takes place for the trial. The mean crossing time over all the trials decreases with the increase of crossing angle. The error-bars indicate the corresponding standard errors of mean.

\paragraph*{S1 Video.}
\label{S1_video}
{\bf Video of the experimental trial which is shown in Fig \ref{real_photo}}

\paragraph*{S2 Video.}
\label{S2_video}
{\bf Edge-cutting process for a trial with $\alpha=89.8\degree$ which is shown in Fig \ref{edge_ex}.} In this video we show the edge-cutting process for a typical trial with $\alpha = 89.8\degree$, the one which has been shown in Fig \ref{edge_ex}. The time frames are shown in scales of $\frac{t-T_i}{T_f-T_i}$.

\paragraph*{S3 Video.}
\label{S3_video}
{\bf Edge-cutting process for a trial with $\alpha=116.9\degree$ which is shown in Fig \ref{edge_ex}.} Same as the previous item with with $\alpha = 116.9\degree$.

\paragraph*{S1 Document.}
\label{S1_Doc}
{\bf Ethical approval statement.}

\section*{Acknowledgments}
The authors gratefully thank A. Sorel, A. Cretual, G. Dachner, and T. Wirth for their help in conducting the experimental trials. A. Marin is also acknowledged for helping in data processing from the camera system.

\section*{Funding Information}
Pratik Mullick thanks financial support from Bretagne S.A.D. as allocated in the project titled ATTRACT. Sylvain Fontaine acknowledges support for his internship by the ``Investissements d’Avenir'' of LabEx PALM (ANR-10-LABX-0039-PALM), in the frame of the PERCEFOULE project.

\section*{Author Contributions}

\hspace{0.47cm}\textbf{Conceptualization:}\\ Pratik Mullick, William H. Warren, Cécile Appert-Rolland, Anne-Hélène Olivier, Julien Pettré

\textbf{Data Curation:}\\Sylvain Fontaine, Anne-Hélène Olivier, Julien Pettré

\textbf{Formal Analysis:}\\Pratik Mullick, William H. Warren, Cécile Appert-Rolland

\textbf{Funding Acquisition:}\\Julien Pettré, Cécile Appert-Rolland

\textbf{Investigation:}\\Pratik Mullick, William H. Warren, Cécile Appert-Rolland, Julien Pettré

\textbf{Methodology:}\\Pratik Mullick, Sylvain Fontaine, William H. Warren, Cécile Appert-Rolland, Julien Pettré

\textbf{Project Administration:}\\Julien Pettré

\textbf{Resources:}\\Julien Pettré

\textbf{Software:}\\Pratik Mullick, Sylvain Fontaine, Anne-Hélène Olivier

\textbf{Supervision:}\\William H. Warren, Cécile Appert-Rolland, Julien Pettré

\textbf{Validation:}\\Pratik Mullick, Sylvain Fontaine, William H. Warren, Cécile Appert-Rolland, Julien Pettré

\textbf{Visualization:}\\Pratik Mullick

\textbf{Writing – Original Draft Preparation:}\\Pratik Mullick

\textbf{Writing – Review \& Editing:}\\Pratik Mullick, William H. Warren, Cécile Appert-Rolland, Julien Pettré

\nolinenumbers

\end{document}